\documentclass[journal]{IEEEtran}
\usepackage{lineno,hyperref}
\modulolinenumbers[5]
\usepackage{amsmath,amssymb,amsfonts}
\usepackage{algorithmic}
\usepackage{graphicx}
\DeclareGraphicsExtensions{.pdf,.jpeg,.png}
\usepackage{subfigure}
\usepackage{epsfig}
\usepackage{booktabs} 
\usepackage{multirow}
\usepackage{array}
\usepackage{dirtree}
\usepackage{epstopdf}
\usepackage{caption,setspace}
\usepackage{wrapfig}
\usepackage{graphics}
\ifCLASSINFOpdf
\else
\fi
\begin{document}

\title{An Improved Reversible Data Hiding in Encrypted Images using Parametric Binary Tree Labeling\\ }
\author{Youqing~Wu,~Youzhi~Xiang,~Yutang~Guo,~Jin~Tang,~and~Zhaoxia~Yin*~\IEEEmembership{Member, IEEE}
\thanks{This research work is supported by the National Natural Science Foundation of China (61872003, U1636206, 61872005, 61860206004), and by Natural Science Foundation of Anhui Higher Education Institutions of China (KJ203A27).}
\thanks{Youqing Wu, Youzhi Xiang, Jin Tang and Zhaoxia Yin are with the school of Key Laboratory of Intelligent Computing and Signal Processing, Ministry of Education, Anhui University, Hefei 230601, P. R. China, e-mail: yinzhaoxia@ahu.edu.cn.}
\thanks{Youqing Wu and Yutang Guo are with the school of Computer Science and Technology, Hefei Normal University, Hefei 230601, P. R. China.}
}

\maketitle

\begin{abstract}
This work proposes an improved reversible data hiding scheme in encrypted images using parametric binary tree labeling(IPBTL-RDHEI), which takes advantage of the spatial correlation in the entire original image but not in small image blocks to reserve room for hiding data. Then the original image is encrypted with an encryption key and the parametric binary tree is used to label encrypted pixels into two different categories. Finally, one of the two categories of encrypted pixels can embed secret information by bit replacement. According to the experimental results, compared with several state-of-the-art methods, the proposed IPBTL-RDHEI method achieves higher embedding rate and outperforms the competitors. Due to the reversibility of IPBTL-RDHEI, the original plaintext image and the secret information can be restored and extracted losslessly and separately.
\end{abstract}


%
\IEEEpeerreviewmaketitle

\section{Introduction}
\setlength{\parskip}{0\baselineskip}
\par Reversible data hiding (RDH) in the plaintext domain is a technique to modify the original cover image to hide secret information (secret data) \cite{Tsai2009Reversible,Chen2013Reversible,Zhang2013Reversible,Li2015Efficient}. It can completely restore the original cover image after extracting the secret information. In the last decade, reversible data hiding has attracted extensive research interest from the information hiding community, due to its potential applications when images are not allowed to be disturbed. As of now, many methods have been designed, which can be mainly classified into three different categories: lossless compression-based \cite{Fridrich2002Lossless,Mehmet2005Lossless}, difference expansion-based \cite{Alattar2004DE,Thodi2007Expansion,Sachnev2009Sorting} and histogram shifting-based \cite{Luo2010HS,Li2013HS} methods. These methods aim to ensure the secret information cannot be detected and the change of the cover image is not perceptible.
\setlength{\parskip}{0\baselineskip}
\par With the increasing demand for user privacy protection on cloud storage, many reversible data hiding schemes in encrypted images (RDHEI) have been published since the pioneering work proposed by Puech \cite{puech2008reversible}. The RDHEI technology embeds secret information into encrypted images rather than plaintext images \cite{Zhou2016Secure,Qian2014Reversible,Ma2013firstRRBE,Xin2017Separable}, which involves three parties: the content-owner, data-hider, and receiver. The original image provider (the content-owner) encrypts the original image before sending it to the cloud. The cloud manager (the data-hider) embeds secret information into encrypted image without knowing the original plaintext image or encryption key. For the receiver, the original plaintext image can be restored and the secret information can be extracted. Fig. 1 shows the framework of RDHEI methods precisely.

\setlength{\parskip}{0\baselineskip}
\par In general, the reported RDHEI techniques can be mainly classified into three different categories, 1) vacating room after encryption (VRAE) \cite{Zhou2016Secure,Zhang2012Separable}; 2) vacating room by encryption (VRBE) \cite{ShuangPBTL}; and 3) reserving room before encryption (RRBE) \cite{1MSB,puyang2018reversible,ccchang}. Since encryption operation disrupts the spatial correlation of the original plaintext image, thus it is difficult for VRAE methods to achieve satisfactory embedding capacity. The VRBE methods use some special encryption schemes to encrypt the original plaintext image while keeping partly spatial correlation in the image after encryption. Since the spatial redundancy is not fully utilized in VRBE methods, thus the embedding capacity is also limited. Different from VRBE and VRAE, the RRBE methods have been proposed to exploit spatial correlation in the original plaintext image, which reserve room before image encryption so as to obtain higher embedding capacity.

\begin{figure}[!ht]
  \centering
    \includegraphics[width=0.45\textwidth]{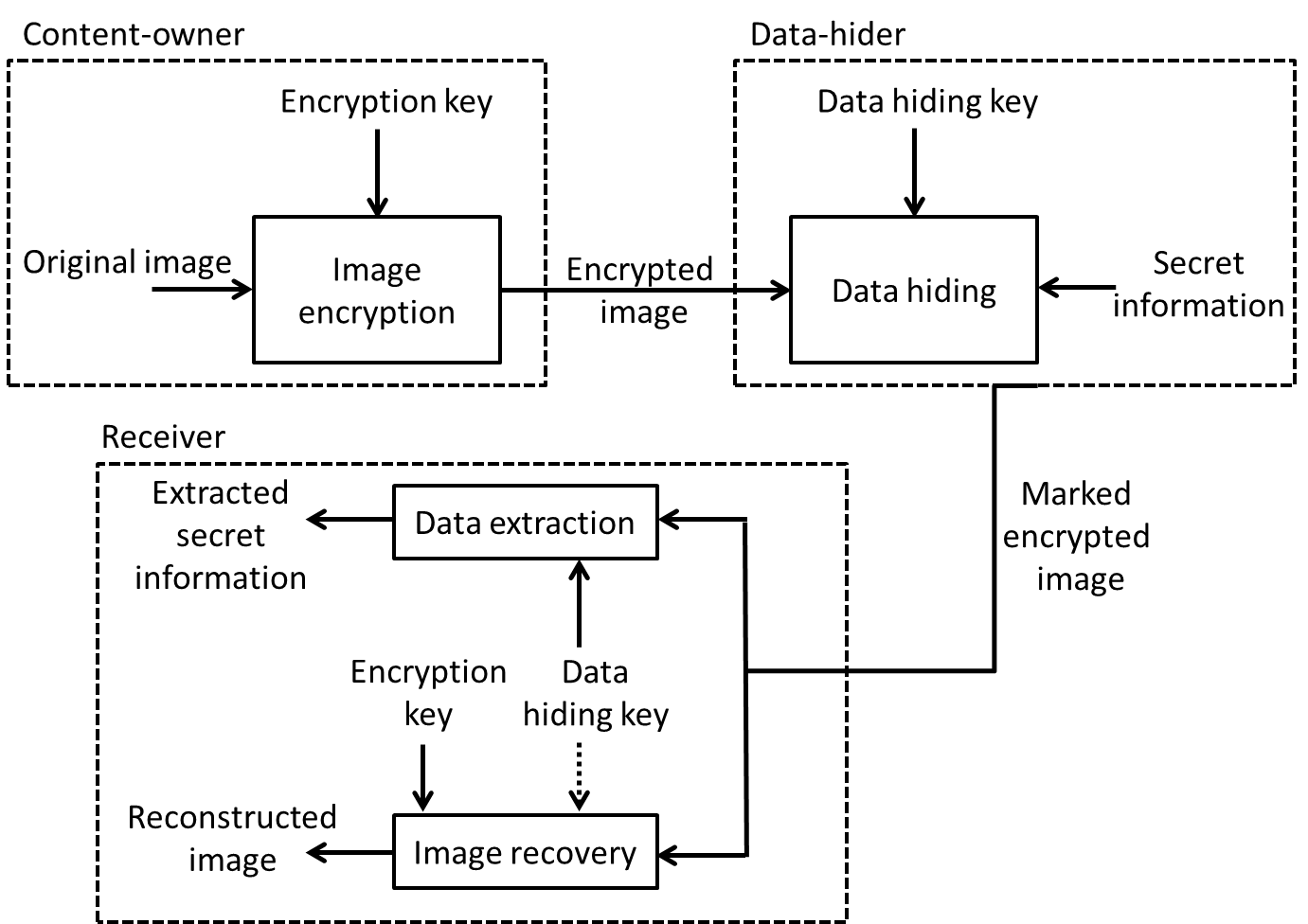}
  \caption{The framework of RDHEI methods.}
\label{fig1}
\end{figure}

\setlength{\parskip}{0\baselineskip}
\par In the previous RDHEI methods, image-recovery and secret information extraction should be processed jointly \cite{Zhou2016Secure}. To separate the process of image-recovery and secret information extraction, separable RDH in the encryption domain have been studied \cite{Zhang2012Separable,ShuangPBTL,1MSB,puyang2018reversible,ccchang}. Zhang \cite{Zhang2012Separable} proposed a separable RDHEI scheme to free up a sparse space to accommodate secret information by compressing the least significant bits. Yi ${ et~al. }$ \cite{ShuangPBTL} proposed a VRBE separable RDHEI method using parametric binary tree labeling to embed secret information by exploiting local correlation within small image blocks. Puteaux ${ et~al. }$ \cite{1MSB} proposed to use MSB substitution to embed secret information. Due to the spatial correlation in a plaintext image, the original image can be restored based on MSB prediction and the secret information can be extracted from the MSB plane. But the method in  \cite{1MSB} only substitutes one-MSB for embedding secret information, thus the embedding rate is lower than one bit per pixel (bpp). Based on \cite{1MSB}, an improved method proposed in \cite{puyang2018reversible} to embed secret information by two-MSB (MSB and second MSB) planes substitution so that the embedding rate can exceed 1 bpp. Chen ${ et~al. }$ \cite{ccchang} transformed the block-based MSB planes of the original plaintext image into bits stream and adopted run-length coding to compress the bits stream for embedding secret information, but the embedding rate is also not very ideal.

\setlength{\parskip}{0\baselineskip}
\par Since Yi ${ et~al. }$ \cite{ShuangPBTL} only used the redundancy in small image blocks but not in the entire image, thus the spatial redundancy is not fully utilized. Based on Yi {${ et~al. }$}'s method \cite{ShuangPBTL}, an improved reversible data hiding scheme in encrypted images using parametric binary tree labeling (IPBTL-RDHEI) is proposed in this paper, which is a high capacity RRBE separable RDHEI method. First, the content-owner reserves embedding room in the plaintext image before encryption and uses a parametric binary tree to label encrypted pixels into two different categories for hiding secret information. Second, the data-hider embeds secret information into one of the two categories of encrypted pixels by bit replacement. Third, according to different permissions, the receiver can obtain the original plaintext image, secret information or both. Compared with Yi {${ et~al. }$}'s method \cite{ShuangPBTL}, the proposed IPBTL-RDHEI method takes full advantage of the image redundancy and achieves a higher embedding rate.

\setlength{\parskip}{0\baselineskip}
\par The main contributions of this paper are as follows:
\setlength{\parskip}{0\baselineskip}
\par 1)	The proposed IPBTL-RDHEI method reserves room in the plaintext image before encryption, which takes full advantage of the spatial correlation in the entire original image but not in small image blocks for embedding data.
\setlength{\parskip}{0\baselineskip}
\par 2)	We present an effective method of RDH in the encryption domain using parametric binary tree labeling and obtain higher embedding rate than state-of-the-art methods. The proposed IPBTL-RDHEI method is separable and error-free in image-recovery and data-extraction.

\setlength{\parskip}{0\baselineskip}
\par The rest of this paper is structured as follows. Section II introduces parametric binary tree labeling scheme. The proposed IPBTL-RDHEI method is elaborated in Section III. Section IV shows the experimental results and analysis. Section V concludes this paper with prospective future works.

\section{Parametric binary tree labeling scheme}
\setlength{\parskip}{0\baselineskip}
\par The pixels in an image can be separated into two different categories by parametric binary tree labeling scheme (PBTL) \cite{ShuangPBTL}, Fig. 2 is a full binary tree that is used to illustrate the distribution of binary codes.

\begin{figure}[!ht]
  \centering
    \includegraphics[width=0.5\textwidth]{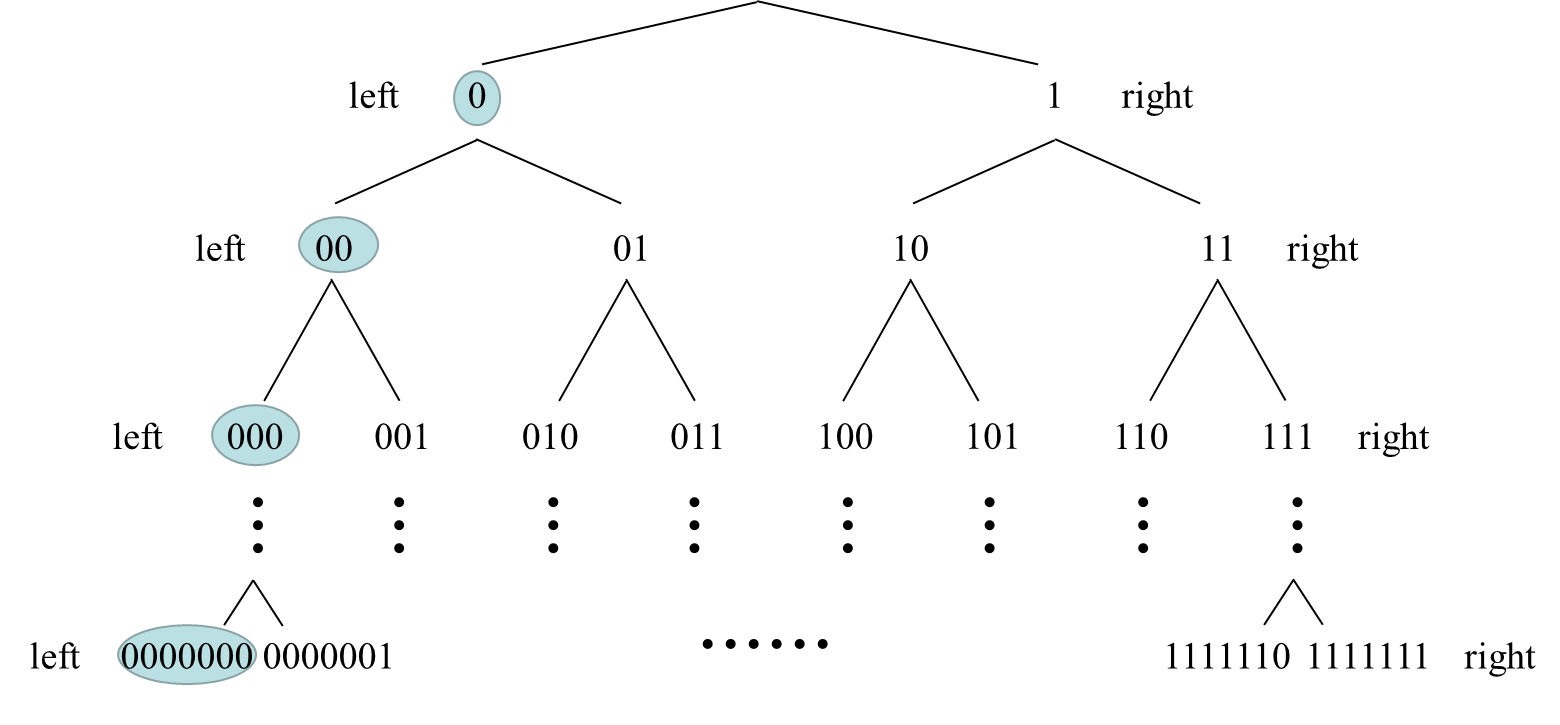}
  \caption{The distribution of binary codes in a full binary tree.}
\label{fig2}
\end{figure}

\setlength{\parskip}{0\baselineskip}
\par For image pixels with 8-bit depth, the full binary tree has 7 layers, and the  ${i^{th}}$ layer has ${2^{i}}$ nodes, where ${i=1,2,...,7}$. Given two parameters ${\alpha}$ and ${\beta}$, where ${1\leqslant \alpha ,\beta\leqslant 7}$, the pixels in two different categories assumed as G1 and G2 are labeled as follows. For G2, all pixels are labeled by the same ${\beta}$-bit of '0...0', which is the first node of the ${\beta ^{th}}$ layer. For G1, all pixels are classified into ${n_{\alpha}}$ different sub-categories, and we use the following Eq. (1) to calculate the positive integer ${n_{\alpha}}$ \cite{ShuangPBTL}:
\begin{equation}
\label{eq::eq1}
\centering
n_{\alpha }= \left\{\begin{matrix}
2^{\alpha }-1 & , & \alpha \leqslant \beta \\
\left ( 2^{\beta }-1 \right )\ast 2^{\alpha -\beta } & , & \alpha > \beta 
\end{matrix}\right.
\end{equation}

\begin{table}[]
\caption{\label{tb::tab1} Illustrative example of labeling bits selection when $\beta$ = 1 and $\alpha$ = 1 to 7.} 
\centering
\setlength{\tabcolsep}{2.8mm}
\begin{tabular}{@{}c|c|c|c@{}}
\toprule
${\beta}$ = 1  & G2 & G1                    & ${n_{\alpha}}$ \\ \midrule
${\alpha}$ = 1 & 0  & 1                     & 1                             \\ \midrule
$\alpha$ = 2 & 0  & 11, 10                & 2                             \\ \midrule
$\alpha$ = 3 & 0  & 111, 110, 101, 100    & 4                             \\ \midrule
$\alpha$ = 4 & 0  & 1111 $\sim$1000       & 8                             \\ \midrule
$\alpha$ = 5 & 0  & 11111 $\sim$10000     & 16                            \\ \midrule
$\alpha$ = 6 & 0  & 111111 $\sim$100000   & 32                            \\ \midrule
$\alpha$ = 7 & 0  & 1111111 $\sim$1000000 & 64                            \\ \bottomrule
\end{tabular}
\end{table}

\begin{table}[]
\caption{\label{tb::tab1} Illustrative example of labeling bits selection when $\beta$ = 2 and $\alpha$ = 1 to 7.}
\centering
\setlength{\tabcolsep}{2.8mm}
\begin{tabular}{@{}c|c|c|c@{}}
\toprule
${\beta}$ = 2  & G2 & G1                           & ${n_{\alpha}}$ \\ \midrule
$\alpha$ = 1 & 00 & 1                            & 1                             \\ \midrule
$\alpha$ = 2 & 00 & 11, 10, 01                   & 3                             \\ \midrule
$\alpha$ = 3 & 00 & 111, 110, 101, 100, 011, 010 & 6                             \\ \midrule
$\alpha$ = 4 & 00 & 1111 $\sim$0100              & 12                            \\ \midrule
$\alpha$ = 5 & 00 & 11111 $\sim$01000            & 24                            \\ \midrule
$\alpha$ = 6 & 00 & 111111 $\sim$010000          & 48                            \\ \midrule
$\alpha$ = 7 & 00 & 1111111 $\sim$0100000        & 96                            \\ \bottomrule
\end{tabular}
\end{table}


\begin{figure*}[!ht]
  \centering
    \includegraphics[width=1\textwidth]{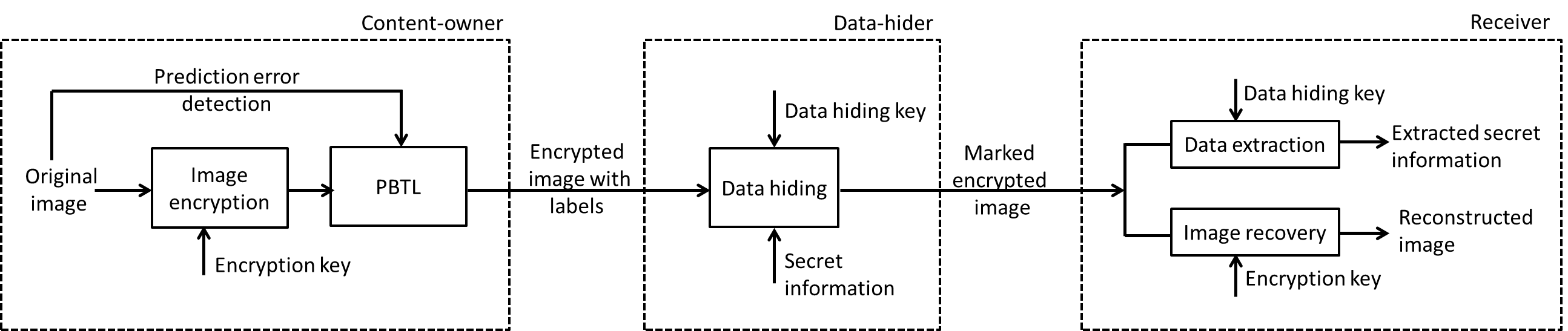}
  \caption{The framework of the proposed IPBTL-RDHEI method.}
\label{fig3}
\end{figure*}

\setlength{\parskip}{0\baselineskip}
\par When ${\alpha \leqslant \beta}$, the ${2^{\alpha }-1}$ nodes from right to left in the  ${\alpha^{th}}$ layer are selected to label ${n_{\alpha}}$ different sub-categories in G1. When ${\alpha > \beta}$, the ${\left ( 2^{\beta }-1 \right )\ast 2^{\alpha -\beta }}$ nodes from right to left in the ${\alpha^{th}}$ layer are selected to label ${n_{\alpha}}$ different sub-categories in G1, that is, when ${\alpha > \beta}$, the selected ${n_{\alpha}}$ binary nodes that are not derived from the first node of the ${\beta ^{th}}$ layer of '0...0'. Moreover, pixels in the same sub-category are labeled with the same ${\alpha}$-bit binary code, and pixels in different sub-categories are labeled with different ${\alpha}$-bit binary codes. Table I and Table II are two illustrative examples of labeling bits selection when ${\beta}$ = 1 to 2 and ${\alpha}$ = 1 to 7.

\setlength{\parskip}{0\baselineskip}
\par As can be seen from Tables I-II, for example, when ${\alpha =3}$, ${\beta =2}$, all the pixels in G2 are labeled by '00', and the ${\left ( 2^{\beta }-1 \right )\ast 2^{\alpha -\beta }=6}$ nodes from right to left in the ${3^{th}}$ layer are selected to label ${6}$ different sub-categories in G1. The ${6}$ selected nodes are '111', '110', '101', '100', '011' and '010', which are not derived from the node of '00', that is, '000' and '001' that derived from '00' are ignored and the remaining nodes in the ${3^{th}}$ layer are kept.

\section{PROPOSED SCHEME}
\setlength{\parskip}{0\baselineskip}
\par The proposed IPBTL-RDHEI method is composed of three main phases: 1) Generation of encrypted image with labels done by the content-owner, 2) Generation of marked encrypted image done by the data-hider, and 3) Data-extraction/image-recovery done by the receiver. In the first phase, the content-owner detects the prediction errors of the original plaintext image and encrypts the original plaintext image using the encryption key. Then, PBTL is used to label encrypted pixels into embeddable pixel set and  non-embeddable pixel set. In the second phase, after using the data-hiding key, the secret information can be hidden by bit replacement in embeddable pixel set. In the third phase, the secret information must be extracted without error from the marked encrypted image with only the data-hiding key, and the original plaintext image must be reconstructed losslessly by exploiting the spatial correlation with only the encryption key. When using both of the keys, the original plaintext image and the secret information must be restored and extracted losslessly. Fig. 3 illustrates the framework of the proposed IPBTL-RDHEI method.

\subsection{Generation of Encrypted Image with labels}
\par This stage has four steps: prediction error detection, image encryption, pixel grouping and pixel labeling using PBTL, which are introduced below:

\begin{figure}[!ht]
  \centering
    \includegraphics[width=0.1\textwidth]{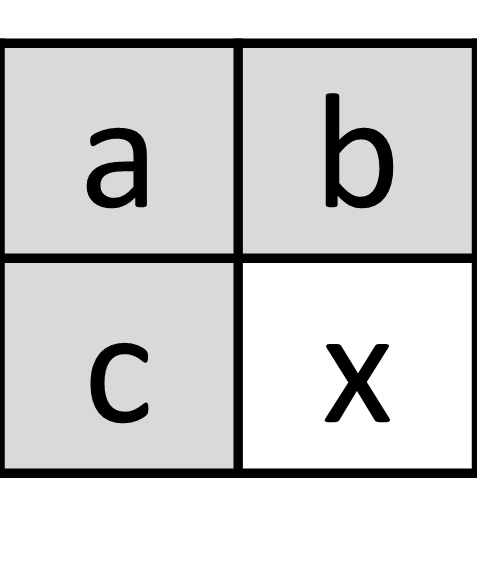}
  \caption{ The context of the MED predictor.}
\label{fig4}
\end{figure}

\subsubsection{Prediction Error Detection}
\par For an original plaintext image, the pixels on the first row and first column are retained as reference pixels. The median edge detector (MED) predictor \cite{Thodi2007Expansion} shown in Fig. 4 can exploit the left, upper and upper left neighboring pixels to predict an image pixel:

\begin{equation}
\label{eq::eq1}
\centering
px = \left\{\begin{matrix}
max(b,c) & , & a\leqslant min(b,c) \\
min(b,c) & , & a\geqslant max(b,c) \\
b+c-a & , & otherwise
\end{matrix}\right.
\end{equation}
where ${px}$ is the prediction value of ${x}$. Hence the prediction error ${e}$ is calculated by:

\begin{equation}
\label{eq::eq1}
\centering
e = x-px
\end{equation}

\subsubsection{Image Encryption}
\par After obtaining all the prediction errors of the 8-bit depth original image ${I}$, we convert each pixel in the original image ${I}$ into 8-bit binary sequence using:

\begin{equation}
\label{eq::eq1}
\centering
x^{k}(i,j)=\left \lfloor x(i,j)/2^{k-1} \right \rfloor mod ~2, k=1,2,...,8
\end{equation}
where ${k}$ is the corresponding bit of the binary sequence, ${1\leqslant i\leqslant m}$ and ${1\leqslant j\leqslant n}$ , ${m\ast n}$ is the size of the original image ${I}$ and ${\left \lfloor * \right \rfloor}$ is floor operation. A pseudo-random matrix ${R}$ of the same size as the original image ${I}$ is generated by an encryption key. Similarly, each pixel ${r(i,j)}$ in ${R}$ is converted into 8-bit binary sequence using Eq. (4). Then the encrypted 8-bit binary sequence can be obtained by the bitwise exclusive-or (XOR) operation:
\begin{equation}
\label{eq::eq1}
\centering
x_{e}^{k}\left ( i,j \right )=x^{k}\left ( i,j \right )\oplus  r^{k}\left ( i,j \right ), k=1,2,...,8
\end{equation}
where ${\oplus}$ is the bitwise XOR operation, and ${x_{e}^{k}\left ( i,j \right )}$ denotes the encrypted 8-bit binary sequence. Finally, Eq. (6) is used to calculate the encrypted pixel ${x_{e}\left ( i,j \right )}$:

\begin{equation}
\label{eq::eq1}
\centering
x_{e}\left ( i,j \right )=\sum_{k=1}^{8}x_{e}^{k}(i,j)\times 2^{k-1}, k=1,2,...,8
\end{equation}

\begin{figure}[!ht]
 \centering
  \subfigure[]{
   \label{fig-5-a}
    \includegraphics[width=0.2\textwidth]{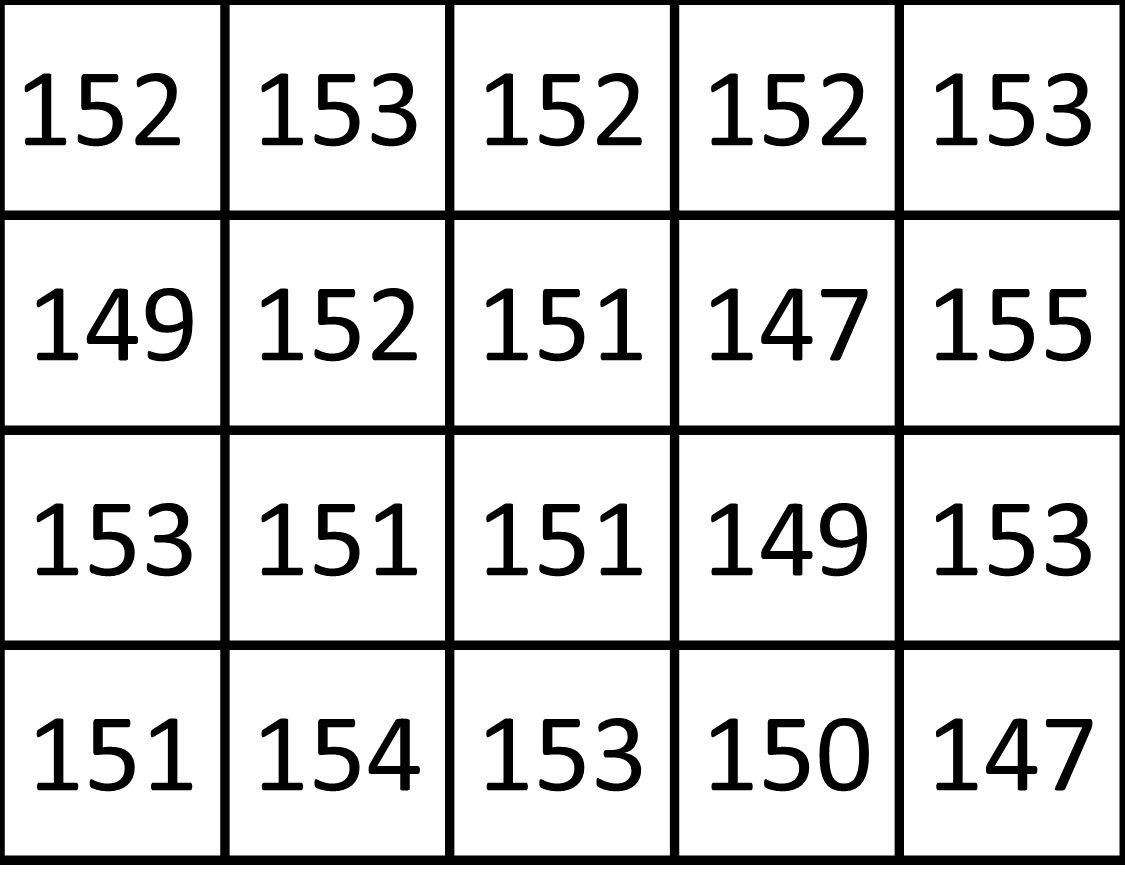}
      }
 \subfigure[]{
   \label{fig-5-b}
    \includegraphics[width=0.2\textwidth]{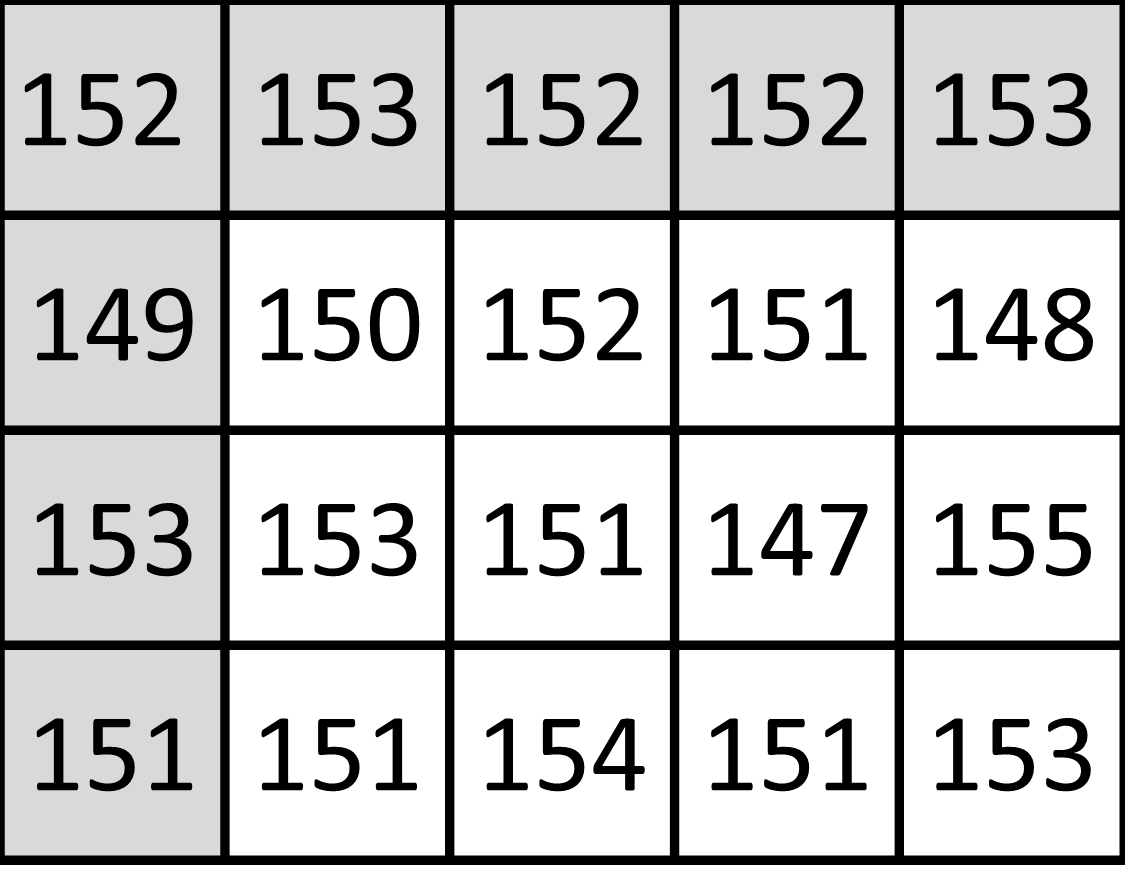} 
    } \\
  \subfigure[]{
   \label{fig-5-c}
    \includegraphics[width=0.2\textwidth]{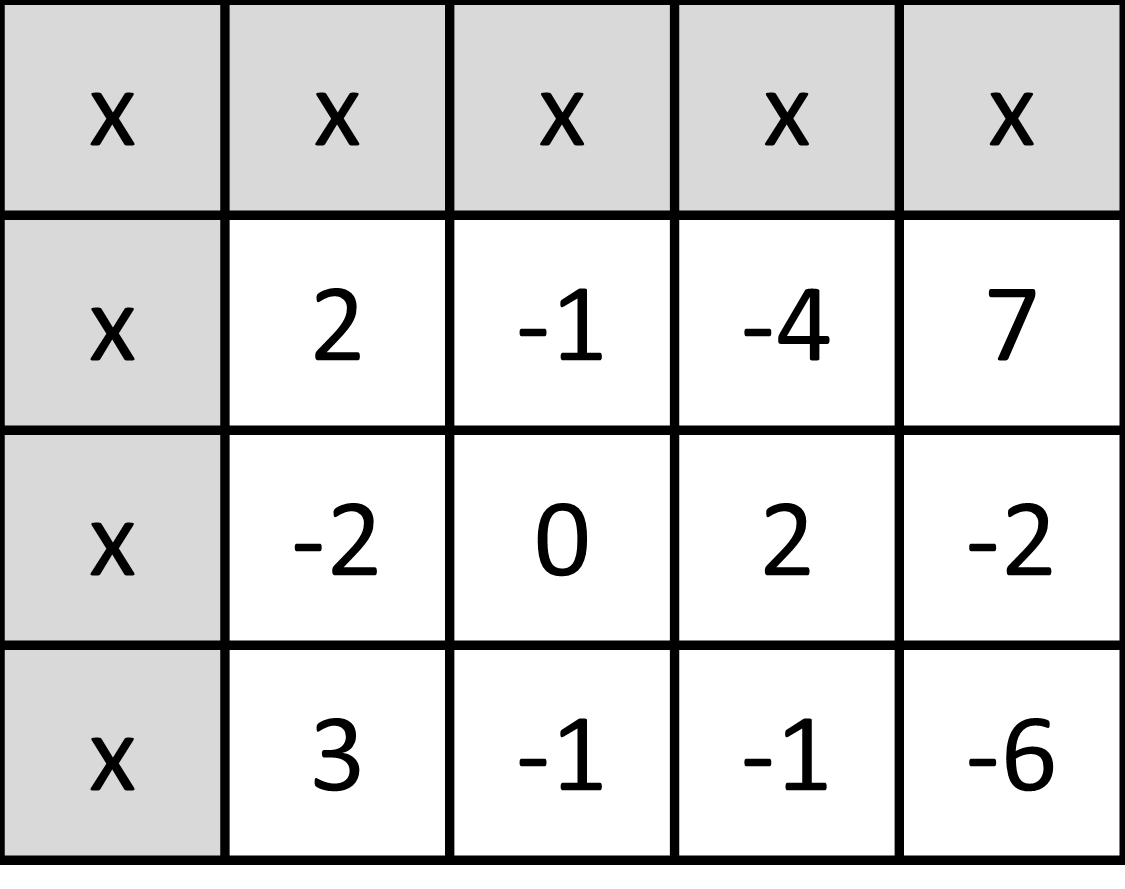} 
    }   
    \subfigure[]{
   \label{fig-5-d}
    \includegraphics[width=0.2\textwidth]{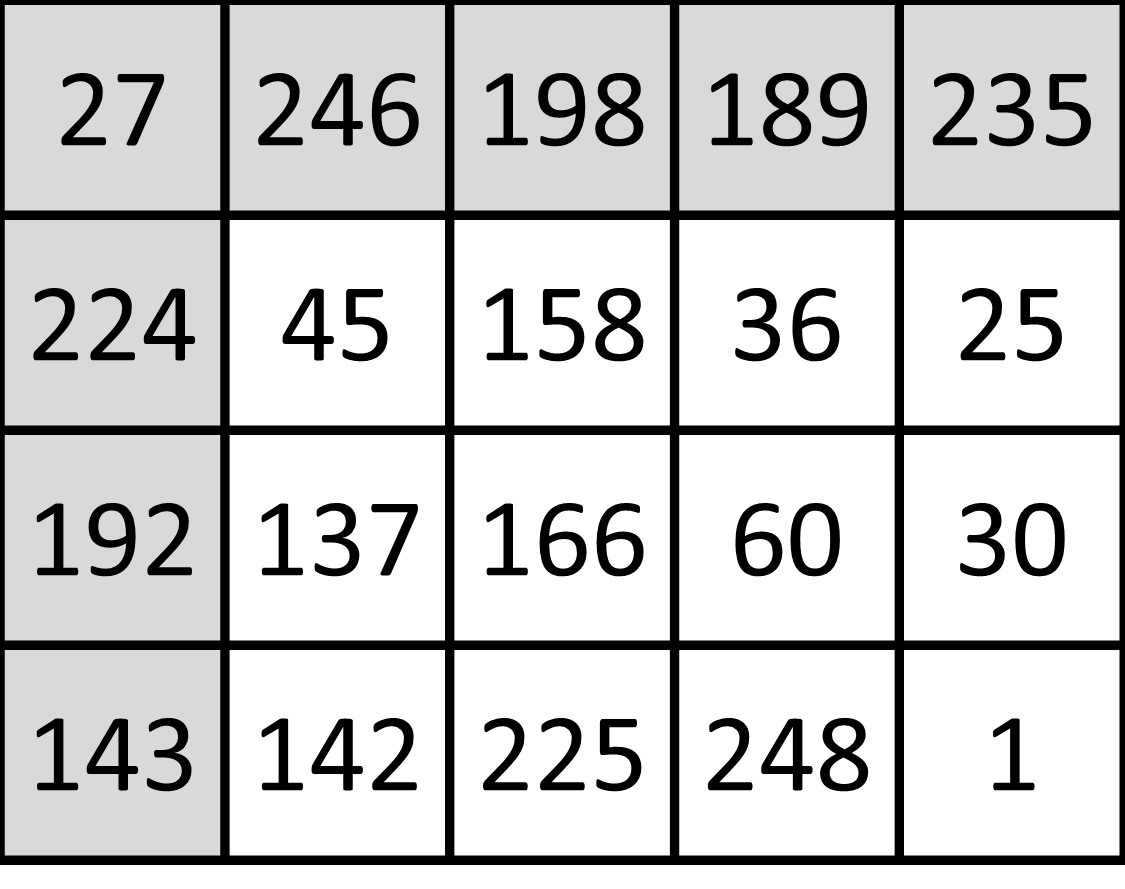} 
    } 
  \caption{Example of prediction error detection and image encryption: (a) original image, (b) prediction values, (c) prediction errors and (d) encrypted image.}
\label{fig5}
\end{figure}

In this way, the encrypted image ${I_{e}}$ is generated. Fig. 5 shows an example of prediction error detection and image encryption. Fig. 5(a) is taken as the original image, where ${m=4}$ and ${n=5}$. The corresponding prediction values of Fig. 5(a) are shown in Fig. 5(b), the pixels on the first row and first column are retained as reference pixels. Fig. 5(c) shows the prediction errors from the subtraction of Fig. 5(a) and Fig. 5(b). Without loss of generality, Fig. 5(d) is assumed to be an encrypted image of Fig. 5(a) by an encryption key ${k_{e}}$.

\subsubsection{Pixel Grouping}
\par We separate all the pixels in encrypted image ${I_{e}}$ into reference pixel set (${P_{r}}$), special pixel set (${P_{s}}$), embeddable pixel set (${P_{e}}$) and non-embeddable pixel set (${P_{n}}$). The pixels on the first row and first column belong to ${P_{r}}$, which will be kept unchanged during the generation of marked encrypted image. We can pick any one pixel to be ${P_{s}}$,  which will be used to store the parameters ${\alpha}$ and ${\beta}$. Then, for each remaining pixel ${I_{e_{i}}\left ( i=1,2,...,m\ast n-(m+n-1)-1 \right )}$, according to the corresponding prediction error ${e_{i}\left ( i=1,2,...,m\ast n-(m+n-1)-1 \right )}$ calculated by Eq. (3), if ${e_{i}}$ meets the condition of Eq. (7) \cite{ShuangPBTL}, the pixel ${I_{e_{i}}}$ belongs to ${P_{e}}$; otherwise, it belongs to ${P_{n}}$. Pixels in ${P_{e}}$ can embed secret information while ${P_{n}}$ cannot.

\begin{equation}
\label{eq::eq1}
\centering
\left \lceil -\frac{n_{\alpha }}{2} \right \rceil\leqslant e_{i}\leqslant \left \lfloor \frac{n_{\alpha }-1}{2} \right \rfloor  
\end{equation}
where ${n_{\alpha}}$ is calculated by Eq. (1), ${\left \lceil * \right \rceil}$ is the ceil operation and ${\left \lfloor * \right \rfloor}$ is the floor operation. Let ${n_{r}}$, ${n_{e}}$ and ${n_{n}}$ represent the number of pixels in ${P_{r}}$, ${P_{e}}$ and ${P_{n}}$, respectively. Thus, ${m*n}$ = ${n_{r} + n_{e} + n_{n} + 1}$, and ${n_{r}}$ = ${ m+n-1}$.

Fig. 6 is the pixel grouping of Fig. 5 when ${\alpha=3}$ and ${\beta=2}$. According to aforementioned, we pick pixels on first row and column as ${P_{r}}$. Without loss of generality, the last pixel is selected as ${P_{s}}$. By the Eq. (1) and Eq. (7). if the prediction error ${e_{i}}$ meets the condition: ${-3\leqslant e_{i}\leqslant 2}$, the pixel ${I_{e_{i}}}$ belongs to ${P_{e}}$; otherwise, it belongs to ${P_{n}}$.
\begin{figure}[!ht]
  \centering
    \includegraphics[width=0.2\textwidth]{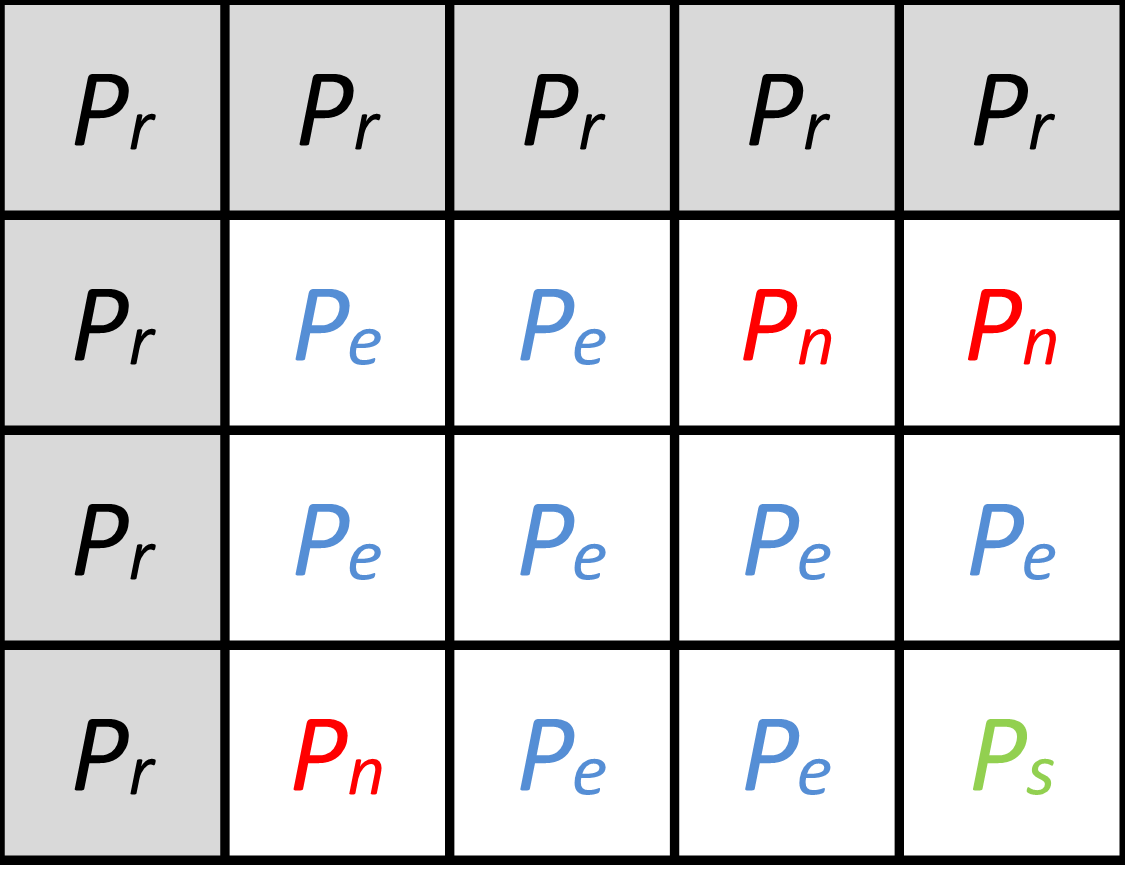}
  \caption{ Pixel grouping.}
\label{fig6}
\end{figure}

\subsubsection{Pixel Labeling using PBTL}
\par Since the positions of ${P_{r}}$ and ${P_{s}}$ are predefined, we just need to label the pixels in ${P_{e}}$ and ${P_{n}}$ using the PBTL scheme. Given two parameters ${\alpha}$ and ${\beta}$, all the pixels in ${P_{n}}$ are labeled by the same ${\beta}$-bit of '0...0', and the remaining ${\left ( 8-\beta \right )}$-bit of each pixel should be kept unchanged. For ${P_{e}}$, all pixels are classified into ${n_{\alpha }}$ different sub-categories according to different prediction errors. Moreover, pixels in the same sub-category are labeled with the same ${\alpha}$-bit binary code, and pixels in different sub-categories are labeled with different ${\alpha}$-bit binary codes. Note that due to the spatial correlation of the original image, the prediction errors of the adjacent pixels are likely to be the same, and then the adjacent pixels are likely to be labeled with the same binary code. If the most significant bits of each pixel are used to be labeled by bit replacement, which may reveal the original image content. To avoid this issue, the least significant bits of each pixel are adopted instead of the most significant bits for labeling, that is, for ${P_{e}}$ and ${P_{n}}$, we arrange the 8-bit binary sequence of each pixel in reverse order before pixel labeling using PBTL.

\subsection{Generation of Marked Encrypted Image}
\par The parameters ${\alpha}$ and ${\beta}$ are first stored in ${P_{s}}$, Since  ${1\leqslant \alpha ,\beta \leqslant 7}$, ${P_{s}}$ is sufficient to store them by bit replacement, then the original 8-bit of ${P_{s}}$ is stored as auxiliary information. In addition, for all the pixels in ${P_{n}}$, the replaced original ${\beta}$-bit of each pixel need to be recorded as auxiliary information. Thus, the auxiliary information contains two parts: the original 8-bit of ${P_{s}}$ and the replaced original ${\beta}$-bit of each pixel in ${P_{n}}$. The payload consists of auxiliary information and secret information.

Each pixel in ${P_{e}}$ is labeled with ${\alpha}$-bit binary code during pixel labeling, then the remaining (8-${\alpha}$)-bit is reserved for hiding payload bits by bit replacement. Therefore, the data-hider can successfully embed the payload of ${\left ( 8-\alpha  \right )\ast n_{e}}$ bits, including auxiliary information of ${8+\beta \ast n_{n}}$ bits and secret information of ${\left ( 8-\alpha  \right )\ast n_{e}-(8+\beta \ast n_{n})}$ bits. For data security, the secret information is first encrypted by using the data hiding key ${k_{d}}$ before the embedding operation. In this way, the marked encrypted image is generated.

\begin{figure}[!ht]
 \centering
  \subfigure[]{
   \label{fig-7-a}
    \includegraphics[width=0.4\textwidth]{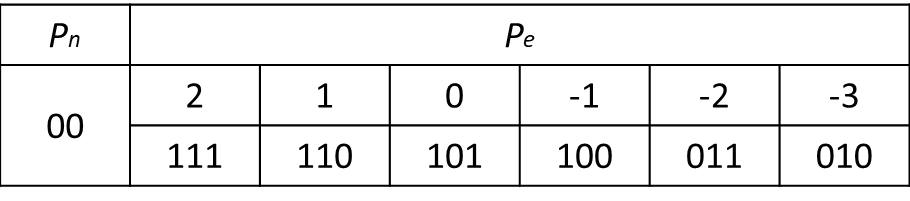}
      }
      
 \subfigure[]{
   \label{fig-7-b}
    \includegraphics[width=0.4\textwidth]{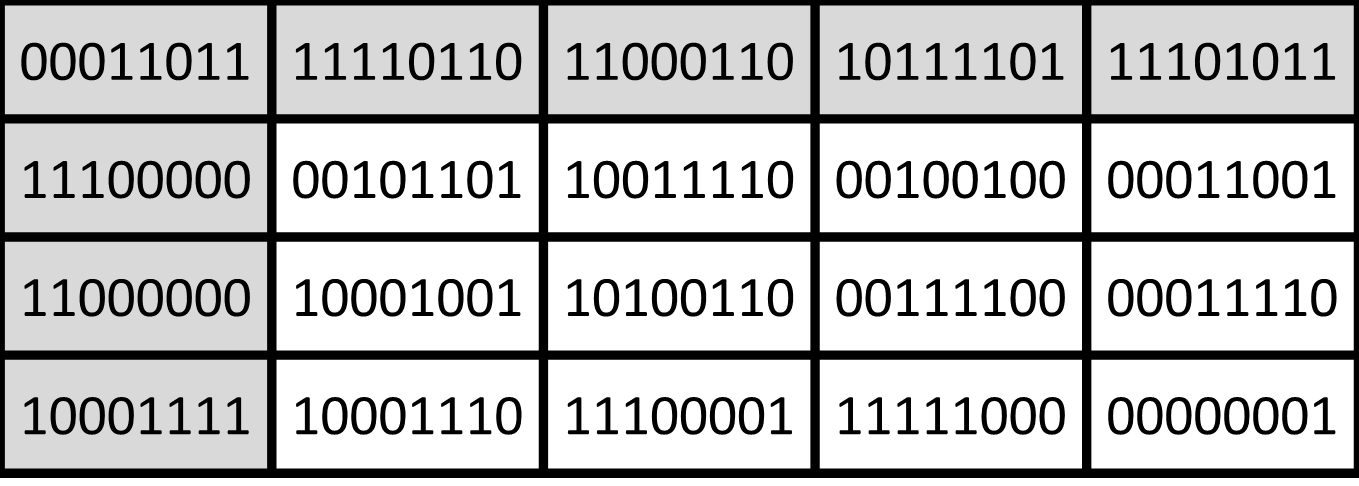} 
    }
    
  \subfigure[]{
   \label{fig-7-c}
    \includegraphics[width=0.4\textwidth]{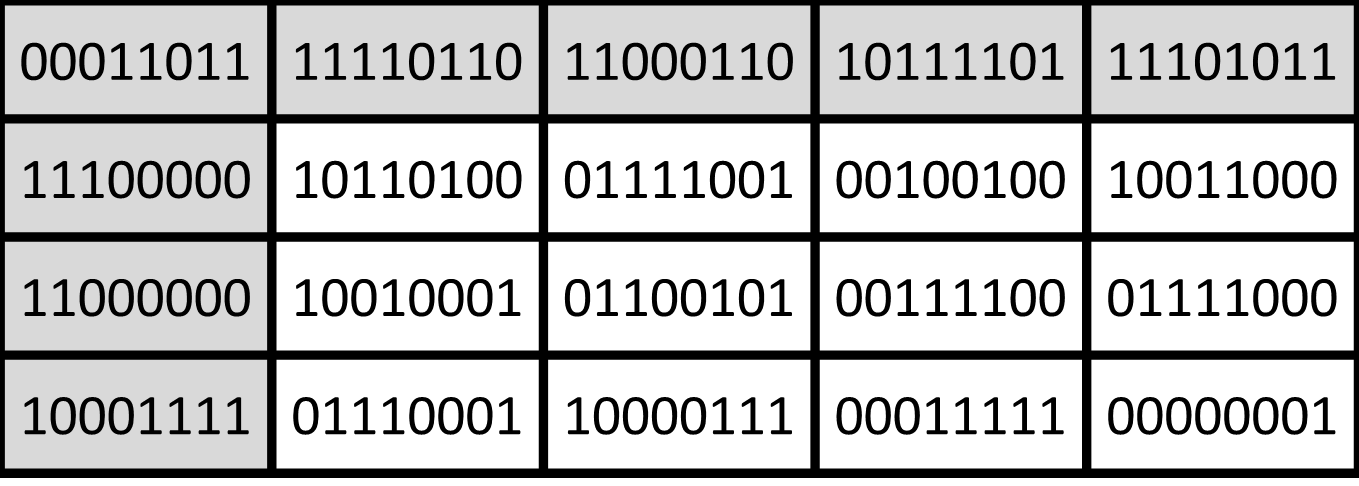} 
    }  
   
\subfigure[]{
   \label{fig-7-d}
    \includegraphics[width=0.4\textwidth]{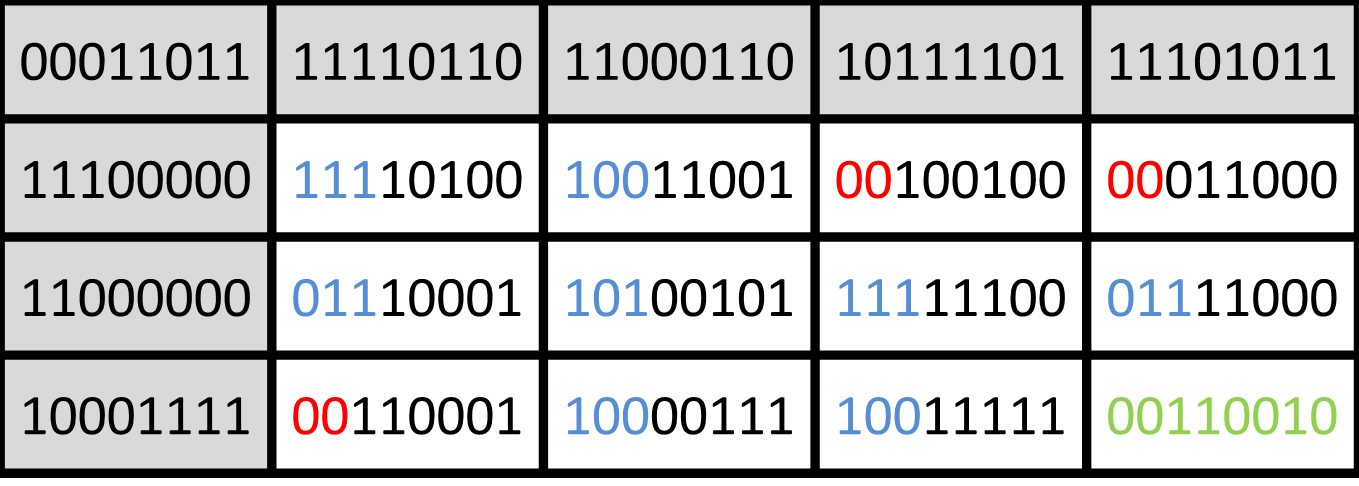} 
    }  
    
\subfigure[]{
   \label{fig-7-e}
    \includegraphics[width=0.4\textwidth]{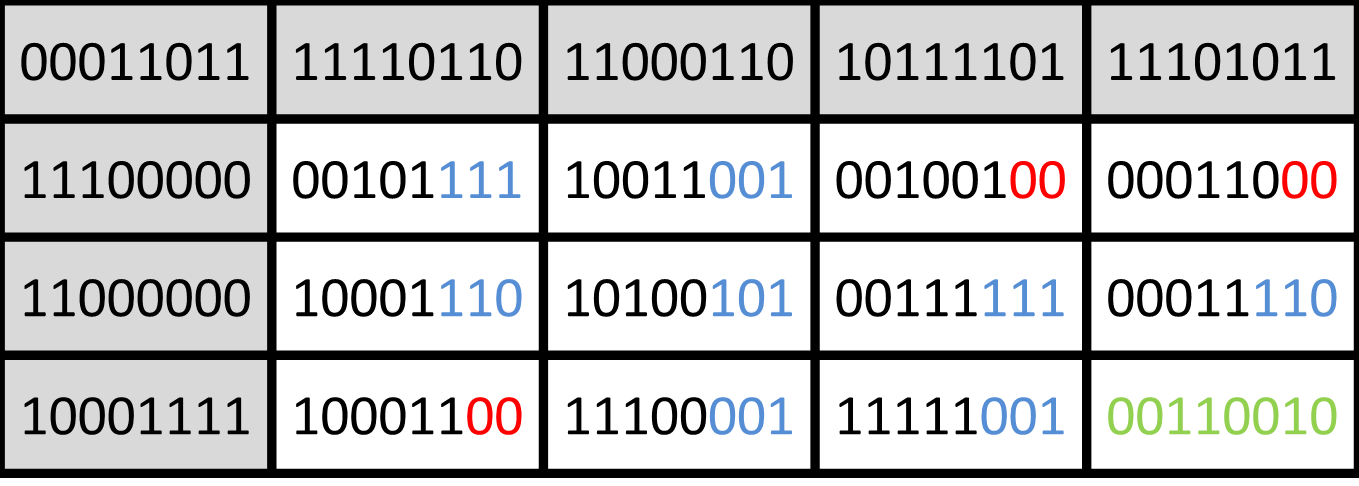} 
    } 
    
 \subfigure[]{
  \label{fig-7-f}
    \includegraphics[width=0.4\textwidth]{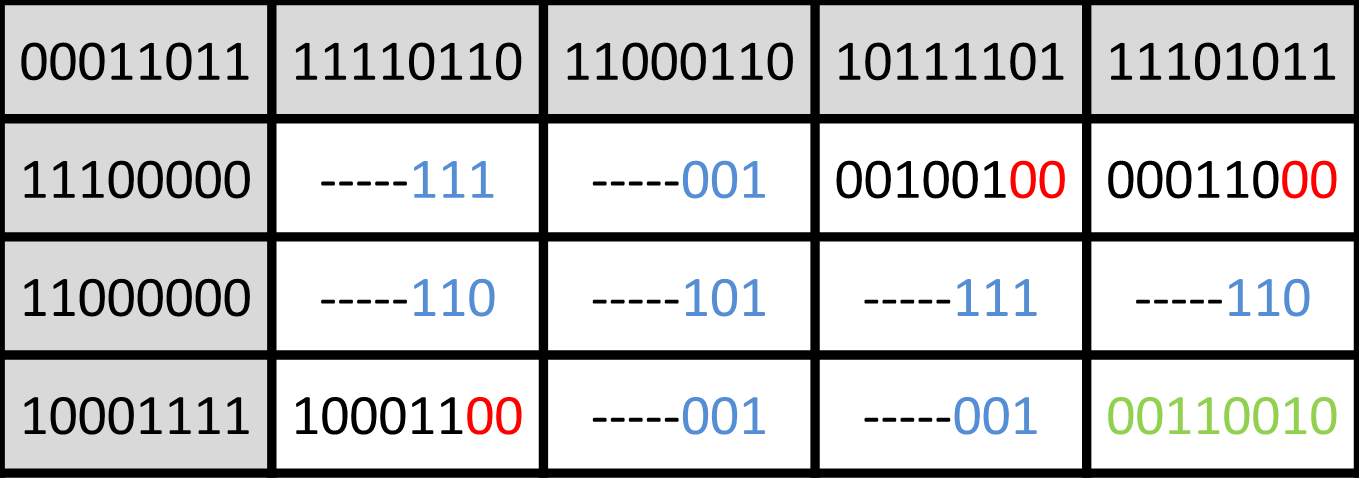} 
   } 
  \caption{Illustrative example of pixel labeling and payload embedding when ${\alpha=3}$ and ${\beta=2}$: (a) Labeling bits selection, (b) The 8-bit binary representation of Fig. 5(d), (c) Reverse order of 8-bit binary sequence in ${P_{e}}$ and ${P_{n}}$, (d) Pixel bits after pixel labeling, (e) Encrypted image with labels and (f) Marked encrypted image.}
\label{fig7}
\end{figure}

\begin{figure*}[!ht]
  \centering
  \subfigure[]{
   \label{fig-8-a}
    \includegraphics[width=0.17\textwidth]{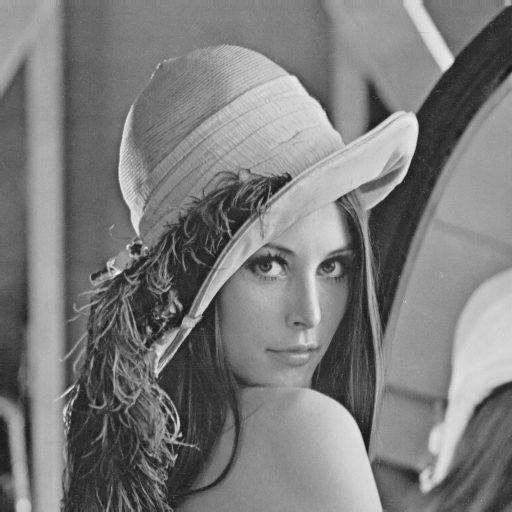}
  }
   \subfigure[]{
   \label{fig-8-b}
    \includegraphics[width=0.17\textwidth]{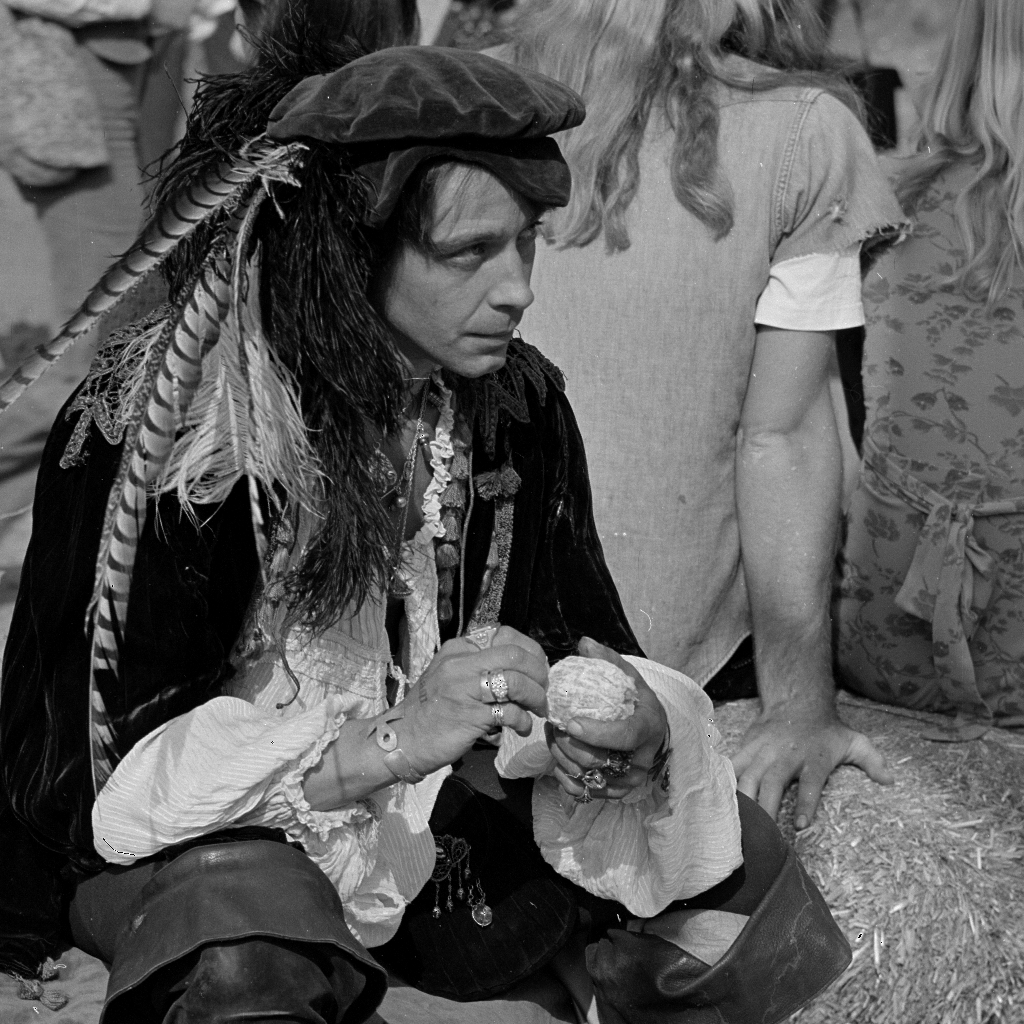}
  }
   \subfigure[]{
   \label{fig-8-c}
    \includegraphics[width=0.17\textwidth]{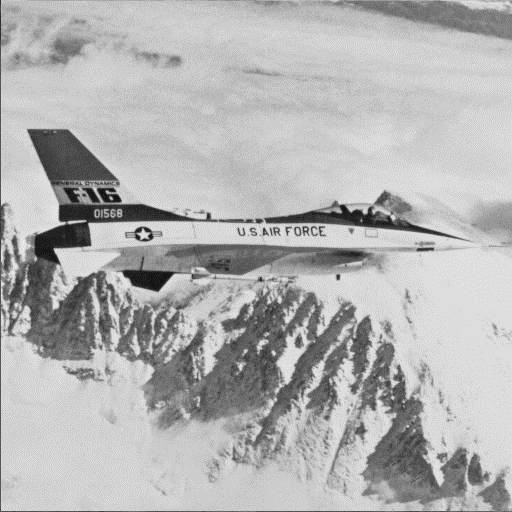}
  }
   \subfigure[]{
   \label{fig-8-d}
    \includegraphics[width=0.17\textwidth]{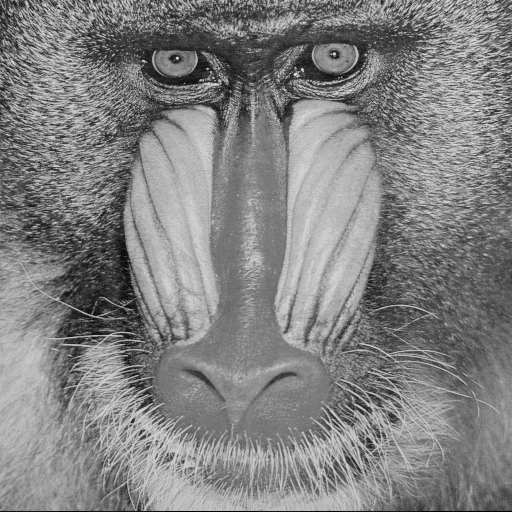}
  }
   \subfigure[]{
   \label{fig-8-e}
    \includegraphics[width=0.17\textwidth]{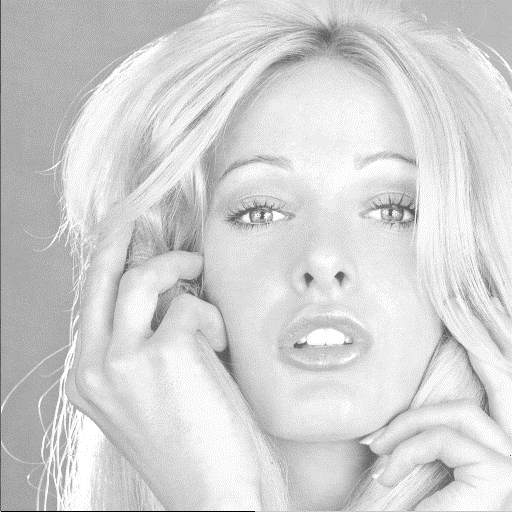}
  }
  \caption{ Test images: (a) ${ Lena }$, (b) ${ Man }$, (c) ${ Jetplane }$, (d) ${ Baboon }$, and (e) ${ Tiffany }$.}
\label{fig8}
\end{figure*}

\begin{figure*}[!ht]
  \centering
  \subfigure[]{
   \label{fig-9-a}
    \includegraphics[width=0.18\textwidth]{Lena.png}
  }
   \subfigure[]{
   \label{fig-9-b}
    \includegraphics[width=0.18\textwidth]{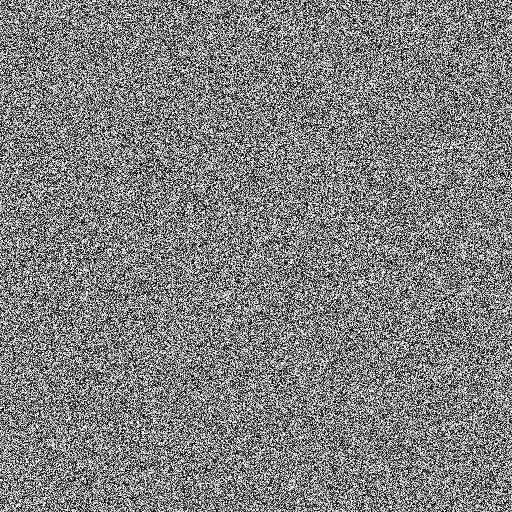}
  }
   \subfigure[]{
   \label{fig-9-c}
    \includegraphics[width=0.18\textwidth]{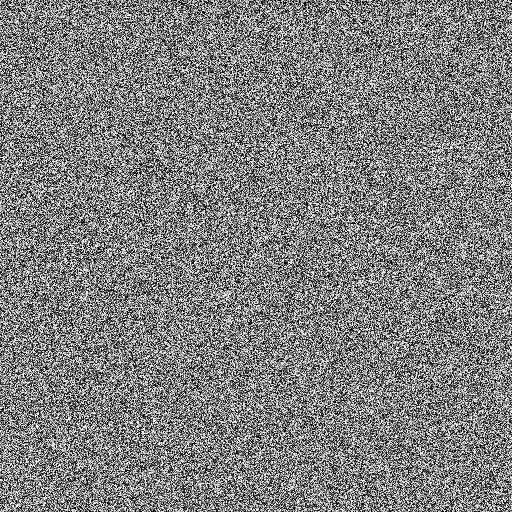}
  }\\
   \subfigure[]{
   \label{fig-9-d}
    \includegraphics[width=0.18\textwidth]{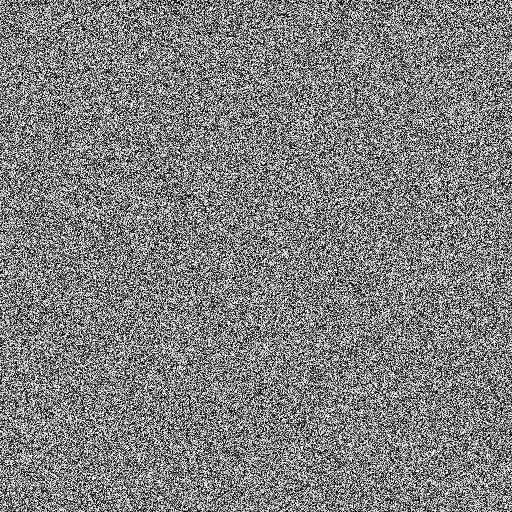}
  }
   \subfigure[]{
   \label{fig-9-e}
    \includegraphics[width=0.18\textwidth]{Lena.png}
  }
   \subfigure[]{
   \label{fig-9-f}
    \includegraphics[width=0.18\textwidth]{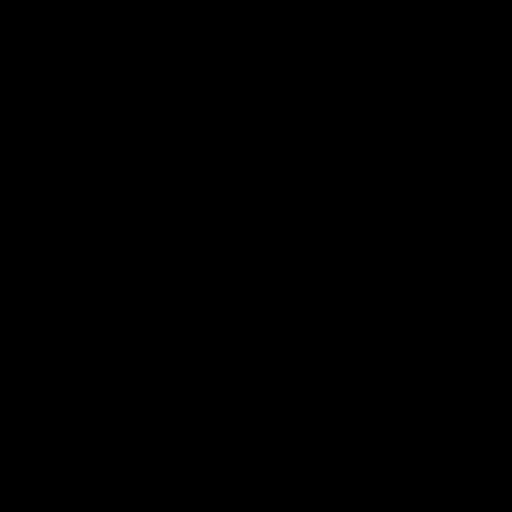}
  }
  \caption{ Simulation results of applying the proposed IPBTL-RDHEI method to ${ Lena }$ image when ${\alpha=5}$ and ${\beta=2}$: (a) original image, (b) encrypted image, (c) encrypted image with labels, (d) marked encrypted image,${r_{max}=2.6447}$ bpp, (e) recovered image, ${PSNR \to +\infty}$ and ${SSIM=1}$,  (f) the difference between (a) and (e)}
\label{fig9}
\end{figure*}

Given different parameters ${\alpha}$ and ${\beta}$, the net embedding rate ${r_{\alpha ,\beta }}$ (bpp) \cite{ShuangPBTL} can be calculated as:

\begin{equation}
\label{eq::eq1}
\centering
r_{\alpha ,\beta }=\frac{(8-\alpha )*n_{e}-(8+\beta *n_{n})}{m*n}
\end{equation}

In practice, we further obtain the maximum net embedding rate ${r_{max}}$ (bpp) \cite{ShuangPBTL} as:
\begin{equation}
\label{eq::eq1}
\centering
r_{max}=max(r_{\alpha ,\beta })_{\alpha =1,\beta =1}^{7}
\end{equation}

Fig. 7 shows an example of pixel labeling and payload embedding when ${\alpha=3}$ and ${\beta=2}$. Fig. 7(a) indicates the labeling bits selection of ${P_{e}}$ and ${P_{n}}$. Here, '00' is used to label each pixel in ${P_{n}}$, '111', '110', '101', '100', '011' and '010' are used to label pixels in ${P_{e}}$ when the prediction error equal to 2, 1, 0, -1, -2 and -3, respectively. Fig. 7(b) represents the 8-bit binary sequence of Fig. 5(d). Fig. 7(c) denotes the reverse order of each pixel in ${P_{e}}$ and ${P_{n}}$. Fig. 7(d) shows pixel bits after pixel labeling. Fig. 7(e) is the encrypted image with labels and Fig. 7(f) is the marked encrypted image after payload embedding. As can be seen, the pixels in ${P_{r}}$ remain unchanged, then the first 4 bits of ${P_{s}}$ are utilized to store ${\alpha}$ and the last 4 bits of ${P_{s}}$ are utilized to store ${\beta}$. Each pixel in ${P_{n}}$ is labeled with '00' and each pixel in ${P_{e}}$ is labeled with 3-bit binary code according to different prediction error. The '- - - - -' in Fig. 7(f) represents the bits that have been embedded payload. Note that the payload contains the auxiliary information of '00000001','00','10' and '01'.

\subsection{Data-Extraction and Image-Recovery}
\par The process of data-Extraction and image-Recovery is the reverse process of the payload embedding. At the receiver end, the secret information must be extracted error-free from the marked encrypted image with only the data-hiding key ${k_{d}}$, and the original plaintext image must be restored losslessly with only the encryption key ${k_{e}}$. According to different permissions, the receiver can obtain the original plaintext image, secret information or both.

\subsubsection{Data-extraction}
\par After obtaining the marked encrypted image, the receiver can extract the secret information. First, we remain the pixels in ${P_{r}}$ unchanged and extract the parameters ${\alpha}$ and ${\beta}$ from ${P_{s}}$. Second, for the rest pixels, we check the labels of their ${\alpha}$ or ${\beta}$ bits in the reverse 8-bit binary sequences and classify them into sets ${P_{e}}$ and ${P_{n}}$. Third, the payload can be extracted from the remaining ${(8-\alpha)}$-bit of each pixel in ${P_{e}}$, then we get the encrypted secret information. Finally, the plaintext secret information can be obtained by decrypting using the data-hiding key ${k_{d}}$.

\subsubsection{Image-recovery}
\par On the other hand, the replaced ${\beta}$-bit of each pixel in ${P_{n}}$ and 8-bit of the pixel in ${P_{s}}$ can be restored using the auxiliary information from the extracted payload. Then the original values of ${P_{n}}$ and ${P_{s}}$ must be obtained by decrypting using the encryption key ${k_{e}}$. Furthermore, the original value of ${P_{r}}$ must be obtained by decrypting directly using the encryption key ${k_{e}}$ as the pixels in ${P_{r}}$ remain unchanged. So far, we restore all the pixels except for the pixels in ${P_{e}}$. For each pixel in ${P_{e}}$, according to its restored left, upper and upper left neighboring pixels, we obtain the corresponding prediction value ${px}$ by Eq. (2), and according to its ${\alpha}$-bit labeling bits, we obtain the corresponding prediction error ${e}$. Then the original value ${x}$ of each pixel in ${P_{e}}$ can be obtained by the following Eq. (10). By now the original content of the image is fully recovered.
\begin{equation}
\label{eq::eq1}
\centering
x = px + e
\end{equation}

Due to the reversibility of each step above, the proposed IPBTL-RDHEI method is separable and error-free in data-extraction and image-recovery.

\section{experimental results and analysis}
\par Several experiments are performed to evaluate the performance of the proposed IPBTL-RDHEI method. Five common 8-bit depth images are used, as shown in Fig. 8. Moreover, in order to reduce the influence caused by the random selection of test images, three datasets including UCID \cite{schaefer2003ucid}, BOSSBase \cite{boss2011}, and BOWS-2 \cite{2017BOWS-2} are also tested respectively. We use two metrics with PSNR (peak signal-to-noise ratio) and SSIM (structural similarity) to evaluate the similarity between two images. The embedding rate (ER) is expressed in bpp and is the key indicator, which is expected to be as large as possible.

\subsection{Performance and Security Analysis}
\par In this section, we evaluate the performance of the proposed IPBTL-RDHEI method on the test images separately. Tables III-V show the maximal embedding rates of test images when ${\beta}$ = 2 to 4 and ${\alpha}$ = 1 to 7. We can see when ${\alpha}$ is small, such as ${\alpha}$ = 1 or 2, the proposed IPBTL-RDHEI method cannot or can only embed a small amount of secret information. The '/' in Tables III-V indicates that the auxiliary information is larger than the reserved room, thus no secret information can be embedded. From Tables III-V, we also can observe that different parameter settings should be selected for different images to reach the maximal embedding rate. In addition, the effect of image texture complexity on embedding rate is significant. A relatively smooth image has a higher embedding rate because there are more pixels belonging to ${P_{e}}$. For example, image ${Jetplane}$ can achieve the maximal embedding rate of 3.0589 bpp when ${\alpha=4}$ and ${\beta=3}$.

\begin{table*}[]
\caption{\label{tb::tab3} Embedding rate ${r_{\alpha ,\beta }}$ (bpp) of test images when $\beta$= 2 and $\alpha$= 1 to 7.}
\centering
\setlength{\tabcolsep}{3mm}
\begin{tabular}{@{}ccllllll@{}}
\toprule
($\alpha$,$\beta$)    & (1,2) & \multicolumn{1}{c}{(2,2)} & \multicolumn{1}{c}{(3,2)} & \multicolumn{1}{c}{(4,2)} & \multicolumn{1}{c}{(5,2)} & \multicolumn{1}{c}{(6,2)} & \multicolumn{1}{c}{(7,2)} \\ \midrule
Lena     & /     & 0.3933                    & 1.6609                    & 2.6867           & 2.6447                    & 1.9285                    & 0.9919                    \\
Man      & /     & \multicolumn{1}{c}{/}     & 0.8173                    & 2.0024                    & 2.4790           & 1.9094                    & 0.9894                    \\
Jetplane & /     & 1.5395                    & 2.6098                    & 3.0250           & 2.6726                    & 1.9223                    & 0.9925                    \\
Baboon   & /     & \multicolumn{1}{c}{/}     & \multicolumn{1}{c}{/}     & 0.2039                    & 0.9692                    & 1.2402           & 0.8615                    \\
Tiffamy  & /     & 0.7108                    & 1.9811                    & 2.8478           & 2.6515                    & 1.9288                    & 0.9928                    \\ \bottomrule
\end{tabular}
\end{table*}
\begin{table*}[]
\caption{\label{tb::tab4} Embedding rate ${r_{\alpha ,\beta }}$ (bpp) of test images when $\beta$= 3 and $\alpha$= 1 to 7.}
\centering
\setlength{\tabcolsep}{3mm}
\begin{tabular}{@{}ccclllll@{}}
\toprule
($\alpha$,$\beta$)    & (1,3) & (2,3)                      & \multicolumn{1}{c}{(3,3)} & \multicolumn{1}{c}{(4,3)} & \multicolumn{1}{c}{(5,3)} & \multicolumn{1}{c}{(6,3)} & \multicolumn{1}{c}{(7,3)} \\ \midrule
Lena     & /     & /                          & 1.7087                    & 2.7872           & 2.6770                    & 1.9407                    & 0.9929                    \\
Man      & /     & /                          & 0.6546                    & 2.0924                    & 2.5517           & 1.9925                    & 0.9915                    \\
Jetplane & /     & \multicolumn{1}{l}{0.9849} & 2.6962                    & 3.0589           & 2.6900                    & 1.9347                    & 0.9939                    \\
Baboon   & /     & /                          & \multicolumn{1}{c}{/}     & \multicolumn{1}{c}{/}     & 0.8789                    & 1.2502           & 0.8896                    \\
Tiffamy  & /     & \multicolumn{1}{l}{0.0525} & 2.0743                    & 2.9204           & 2.6793                    & 1.9416                    & 0.9946                    \\ \bottomrule
\end{tabular}
\end{table*}
\begin{table*}[]
\caption{\label{tb::tab5} Embedding rate ${r_{\alpha ,\beta }}$ (bpp) of test images when $\beta$= 4 and $\alpha$= 1 to 7.}
\centering
\setlength{\tabcolsep}{3mm}
\begin{tabular}{@{}ccclllll@{}}
\toprule
($\alpha$,$\beta$)    & (1,4) & (2,4)                      & \multicolumn{1}{c}{(3,4)} & \multicolumn{1}{c}{(4,4)} & \multicolumn{1}{c}{(5,4)} & \multicolumn{1}{c}{(6,4)} & \multicolumn{1}{c}{(7,4)} \\ \midrule
Lena     & /     & /                          & 1.2997                    & 2.7703           & 2.6693                    & 1.9423                    & 0.9933                    \\
Man      & /     & /                          & 0.1126                    & 2.0374                    & 2.5516           & 1.9226                    & 0.9919                    \\
Jetplane & /     & \multicolumn{1}{l}{0.4303} & 2.4107                    & 3.0266           & 2.6778                    & 1.9360                    & 0.9942                    \\
Baboon   & /     & /                          & \multicolumn{1}{c}{/}     & \multicolumn{1}{c}{/}     & 0.6866                    & 1.2002           & 0.8953                    \\
Tiffamy  & /     & /     & \multicolumn{1}{l}{1.7111}        & 2.8941                           & 2.6703                    & 1.9436                    & 0.9949                    \\ \bottomrule
\end{tabular}
\end{table*}

\setlength{\parskip}{0\baselineskip}
\par Fig. 9 takes ${ Lena }$ as an example to show different images in different phases generated by the proposed IPBTL-RDHEI method. Fig. 9(a) is the original image. Fig. 9(b) shows the encrypted image obtained by an encryption key ${ k_{e} }$. The encrypted image with labels is shown in Fig. 9(c). Fig. 9(d) presents the marked encrypted image. Fig. 9(e) gives the recovered image, which is the same as Fig. 9(a). Fig. 9(f) is the difference between Fig. 9(a) and Fig. 9(e), where all pixels are 0. Fig. 9(b), (c) and (d) are three encrypted versions of Fig. 9(a), and it is difficult to detect the content of Fig. 9(a) from Fig. 9(b), (c) and (d), which means that the proposed IPBTL-RDHEI method has a high perceptual security level.

\setlength{\parskip}{0\baselineskip}
\par  To further test the security of our method, Tables VI-VIII show the PSNR and SSIM values for each encrypted version image with the corresponding original image. From Tables VI-VIII, we can see that the PSNR value of each encrypted version image is very low and the SSIM value of each encrypted version image is almost 0. Thus no information can be obtained from these encrypted version images, which means that the proposed IPBTL-RDHEI method securely protects the privacy of the original image and can be applied to the RDH in the encryption domain.

\begin{table}[]
\caption{\label{tb::tab6} Encrypted images' PSNR and SSIM  with the original images when ${\alpha=5}$ and ${\beta=2}$.}
\setlength{\tabcolsep}{0.7mm}
\begin{tabular}{@{}cccccc@{}}
\toprule
Encrypted image & Lena   & Man    & Jetplane & Baboon & Tiffany \\ \midrule
PSNR (dB)            & 9.2255 & 7.9937 & 8.0077   & 9.5108 & 6.8839  \\
SSIM            & 0.0341 & 0.0681 & 0.0346   & 0.0299 & 0.0389  \\ \bottomrule
\end{tabular}
\end{table}
\begin{table}[]
\caption{\label{tb::tab7} Encrypted images with labels' PSNR and SSIM  with the original images when ${\alpha=5}$ and ${\beta=2}$.}
\setlength{\tabcolsep}{0.7mm}
\begin{tabular}{@{}cccccc@{}}
\toprule
Encrypted image  & Lena   & Man    & Jetplane & Baboon & Tiffany \\
with labels      &        &        &          &        &         \\
\midrule
PSNR (dB)             & 9.2256 & 8.0096 & 7.9935   & 9.5215 & 6.8741  \\
SSIM            & 0.0347 & 0.0695 & 0.0353   & 0.0306 & 0.0391  \\ \bottomrule
\end{tabular}
\end{table}
\begin{table}[]
\caption{\label{tb::tab8} Marked encrypted images' PSNR and SSIM  with the original images when ${\alpha=5}$ and ${\beta=2}$.}
\setlength{\tabcolsep}{0.7mm}
\begin{tabular}{@{}cccccc@{}}
\toprule
Marked encrypted  & Lena   & Man    & Jetplane & Baboon & Tiffany \\ 
image             &        &        &          &        &         \\
\midrule
PSNR (dB)                     & 9.2222 & 8.0176 & 7.9941   & 9.5182 & 6.8838  \\
SSIM                    & 0.0351 & 0.0694 & 0.0359   & 0.0325 & 0.0375  \\ \bottomrule
\end{tabular}
\end{table}

\begin{figure*}[!ht]
  \centering
    \includegraphics[width=0.6\textwidth]{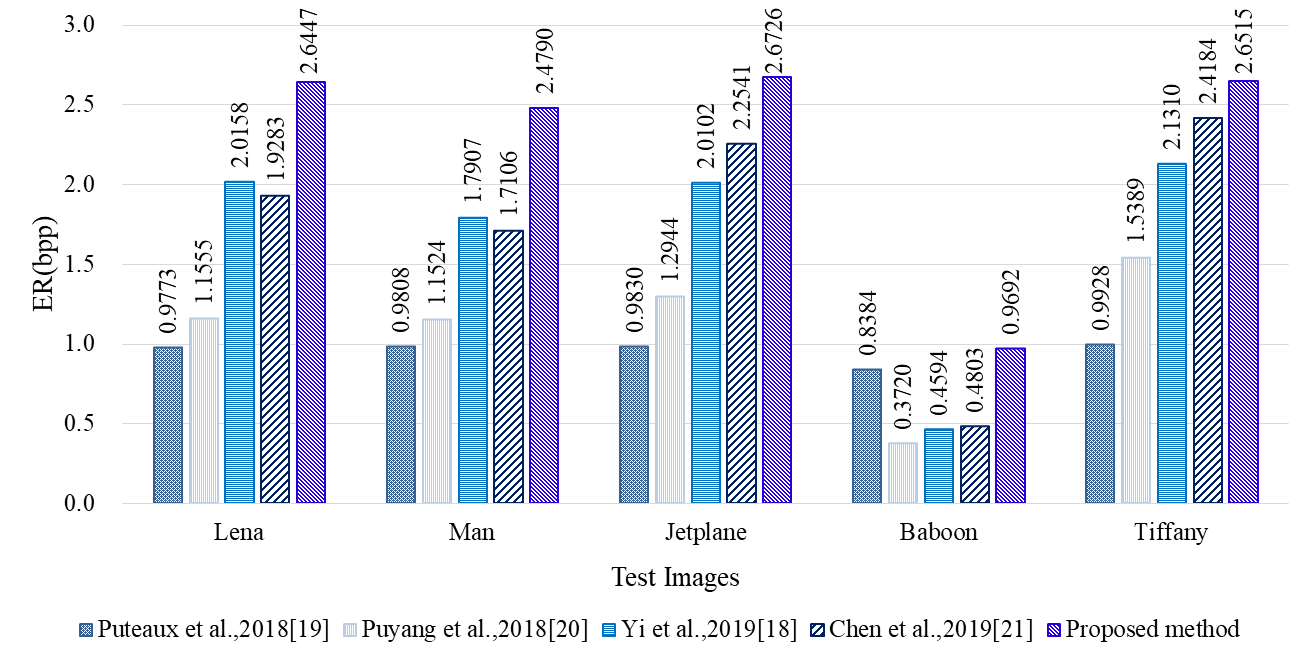}
  \caption{ Comparison of maximal embedding rates of test images between our method and four state-of-the-art methods.}
\label{fig10}
\end{figure*}

\begin{figure*}[!ht]
  \centering
    \includegraphics[width=0.6\textwidth]{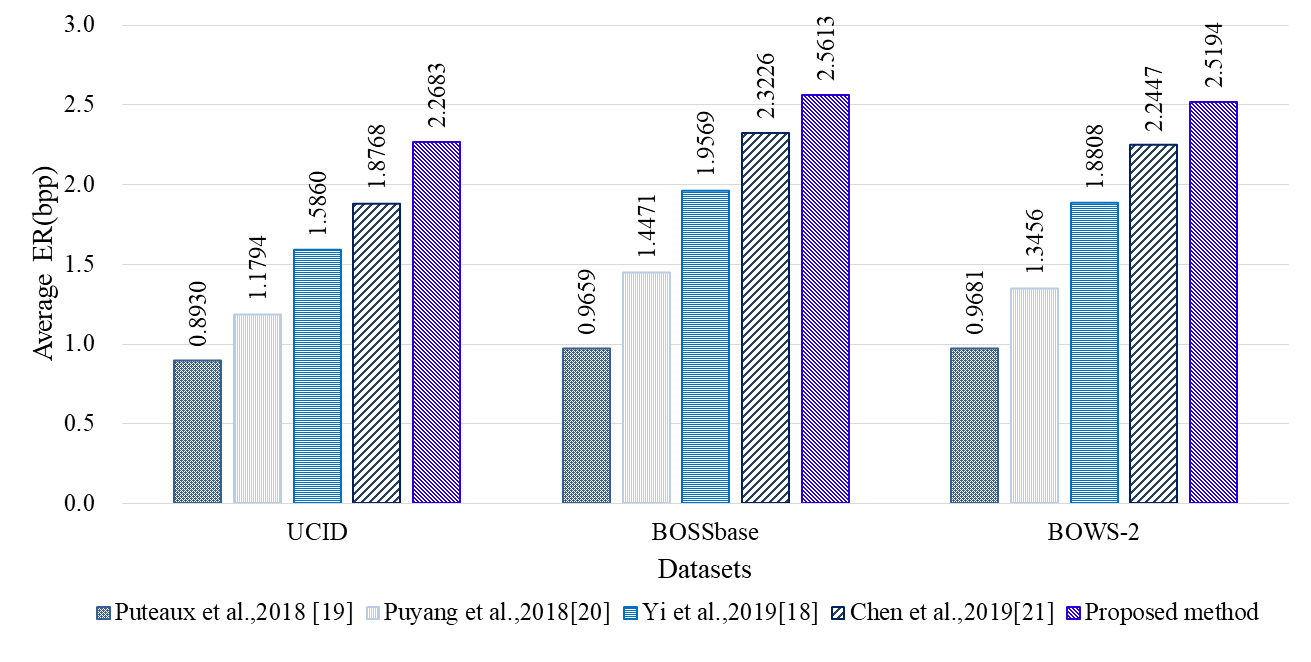}
  \caption{ Comparison of the average embedding rates of three datasets between our method and four state-of-the-art methods.}
\label{fig11}
\end{figure*}

\subsection{Comparisons with Related Methods and Analysis}
\par In this section, we compare the embedding rate of the proposed IPBTL-RDHEI method with several state-of-the-art methods. The parameters ${\alpha}$ and ${\beta}$ in the proposed IPBTL-RDHEI method are set to 5 and 2. To obtain a better performance, we set the length of fixed-length codewords to 3 and block size to ${4\times 4}$ in \cite{ccchang}. In \cite{ShuangPBTL}, the parameters ${\alpha}$ and ${\beta}$ are also set to 5 and 2, and the block size is set to ${3\times 3}$. 

\setlength{\parskip}{0\baselineskip}
\par Fig. 10 shows the maximal embedding rates of test images, compared with four competitors  \cite{ShuangPBTL}, \cite{1MSB}, \cite{puyang2018reversible} and \cite{ccchang}. We can see that the proposed IPBTL-RDHEI method achieves higher embedding rate and outperforms the competitors.

\begin{table}[]
\caption{\label{tb::tab9} Detailed embedding rates of our method on the three datasets when $\alpha$= 5 and $\beta$= 2 .}
\centering
\setlength{\tabcolsep}{0.9mm}
\begin{tabular}{@{}lcccc@{}}
\toprule
Datasets                                  & \multicolumn{1}{l}{Indicators} & \multicolumn{1}{l}{Best case} & \multicolumn{1}{l}{Worst case} & \multicolumn{1}{l}{Average} \\ \midrule
\multirow{3}{*}{UCID}                 & ER (bpp)                             & 2.9759                        & 0                              & 2.2683                      \\ \cmidrule(l){2-5} 
                                          & PSNR (dB)                           & +${ \infty }$        & +${ \infty }$         & +${ \infty }$      \\ \cmidrule(l){2-5} 
                                          & SSIM                           & 1                             & 1                              & 1                           \\ \midrule
\multirow{3}{*}{BOSSbase}                   & ER (bpp)                             & 2.9883                        & 0.0713                         & 2.5613                      \\ \cmidrule(l){2-5} 
                                          & PSNR (dB)                          & +${ \infty }$        & +${ \infty }$      & +${ \infty }$     \\ \cmidrule(l){2-5} 
                                          & SSIM                           & 1                             & 1                              & 1                           \\ \midrule
\multirow{3}{*}{BOWS-2} & ER (bpp)                             & 2.9883                        & 0.0484                         & 2.5194                      \\ \cmidrule(l){2-5} 
\multicolumn{1}{c}{}                      & PSNR (dB)                          & +${ \infty }$        & +${ \infty }$         & +${ \infty }$      \\ \cmidrule(l){2-5} 
\multicolumn{1}{c}{}                      & SSIM                           & 1                             & 1                              & 1                           \\ \bottomrule
\end{tabular}
\end{table}

\setlength{\parskip}{0\baselineskip}
\par Moreover, in order to reduce the influence caused by the random selection of test images, the detailed embedding rates of the proposed IPBTL-RDHEI method on the three datasets are shown in Table IX. For the best cases, the embedding rates are 2.9759 bpp, 2.9883 bpp, and 2.9883 bpp, respectively. Since ${\alpha}$ is set to 5, that is, each pixel in ${P_{e}}$ is labeled with 5 bits, and the remaining 3 bits can be embedded payload bits by bit replacement, thus the embedding rates approach 3 bpp in the best cases. In the UCID dataset, the worst embedding rate is 0 bpp, which means that the auxiliary information is larger than the reserved room, therefore no secret information is embedded when ${\alpha=5}$ and ${\beta=2}$. Also, Table IX indicates that each original plaintext image can be recovered error-free (${PSNR \to  + \infty }$ and ${SSIM = 1}$).

\setlength{\parskip}{0\baselineskip}
\par Fig. 11 compares the average embedding rates of the three datasets between the proposed IPBTL-RDHEI method and these four state-of-the-art methods. The average embedding rates on the three datasets are close to 1 bpp but no more than 1 bpp in the EPE-HCRDH method \cite{1MSB}. The method of two-MSB planes substitution in  \cite{puyang2018reversible} has higher embedding rate than EPE-HCRDH \cite{1MSB}. In addition, the average embedding rates of Chen {${ et~al. }$}'s method \cite{ccchang} are higher, reaching  1.8768 bpp in the UCID dataset, 2.3226 bpp in the BOSSBase dataset and 2.2447 bpp in the BOWS-2 dataset, respectively. Both Yi {${ et~al. }$}'s method \cite{ShuangPBTL} and our method are based on PBTL. The results in  Fig. 11 show that the proposed IPBTL-RDHEI method significantly improves the embedding rate compared with Yi {${ et~al. }$}'s method \cite{ShuangPBTL}. There are two main reasons for this: first, the proposed IPBTL-RDHEI method reserves room in the plaintext image before encryption, which can take full advantage of the image redundancy; second, we take advantage of the spatial correlation in the entire original image but not in small image blocks to reserve room for embedding data, which reduces the number of ${P_{r}}$, that results in less auxiliary information. Based on the above analysis, we can see that the proposed IPBTL-RDHEI method has better performance.

\section{Conclusion}
\setlength{\parskip}{0\baselineskip}
\par This paper presents an effective method of RDH in the encryption domain using parametric binary tree labeling, which is an improved method based on Yi {${ et~al. }$}'s work \cite{ShuangPBTL}. The proposed IPBTL-RDHEI method provides a good level of security that can be applied to protect the privacy of the original plaintext image. In addition, compared with the state-of-the-art methods, the proposed IPBTL-RDHEI method takes full advantage of the image redundancy, which not only is separable and error-free in image-recovery and data-extraction but also achieves higher embedding rate. In further research, we will test other error predictors to make more pixels into the embeddable pixel set, then more pixels can be utilized to embed secret information.



\bibliographystyle{unsrt}
\bibliography{name}

\begin{thebibliography}{10}

\bibitem{Tsai2009Reversible}
Piyu Tsai, YuChen Hu, and Hsiu~Lien Yeh.
\newblock Reversible image hiding scheme using predictive coding and histogram
  shifting.
\newblock {\em Signal Processing}, 89(6):1129--1143, 2009.

\bibitem{Chen2013Reversible}
Xianyi Chen, Xingming Sun, Huiyu Sun, Zhili Zhou, and Jianjun Zhang.
\newblock Reversible watermarking method based on asymmetric-histogram shifting
  of prediction errors.
\newblock {\em The Journal of Systems and Software}, 86(10):2620--2626, 2013.

\bibitem{Zhang2013Reversible}
Xinpeng Zhang.
\newblock Reversible data hiding with optimal value transfer.
\newblock {\em IEEE Transactions on Multimedia}, 15(2):316--325, 2013.

\bibitem{Li2015Efficient}
Xiaolong Li, Weiming Zhang, Xinlu Gui, and Bin Yang.
\newblock Efficient reversible data hiding based on multiple histograms
  modification.
\newblock {\em IEEE Transactions on Information Forensics and Security},
  10(9):2016--2027, 2015.

\bibitem{Fridrich2002Lossless}
Jessica Fridrich, Miroslav Goljan, and Rui Du.
\newblock Lossless data embedding: New paradigm in digital watermarking.
\newblock {\em EURASIP J. Appl. Signal Process}, 2002(2):185--196, 2002.

\bibitem{Mehmet2005Lossless}
Mehmet~Utku Celik, Gaurav Sharma, Ahmet~Murat Tekalp, and Eli Saber.
\newblock Lossless generalized-lsb data embedding.
\newblock {\em IEEE Transactions on Image Processing}, 14(2):253--266, 2005.

\bibitem{Alattar2004DE}
Adnan~M Alattar.
\newblock Reversible watermark using the difference expansion of a generalized
  integer transform.
\newblock {\em IEEE Transactions on Image Processing}, 13(8):1147--1156, 2004.

\bibitem{Thodi2007Expansion}
Diljith~M Thodi and Jeffrey~J Rodríguez.
\newblock Expansion embedding techniques for reversible watermarking.
\newblock {\em IEEE Transactions on Image Processing}, 16(3):721--730, 2007.

\bibitem{Sachnev2009Sorting}
Vasiliy Sachnev, Hyoung~Joong Kim, Jeho Nam, Sundaram Suresh, and Yunqing Shi.
\newblock Reversible watermarking algorithm using sorting and prediction.
\newblock {\em IEEE Transactions on Circuits and Systems for Video Technology},
  19(7):989--999, 2009.

\bibitem{Luo2010HS}
Lixin Luo, Zhenyong Chen, Ming Chen, Xiao Zeng, and Zhang Xiong.
\newblock Reversible image watermarking using interpolation technique.
\newblock {\em IEEE Transactions on Information Forensics and Security},
  5(1):187--193, 2010.

\bibitem{Li2013HS}
Xiaolong Li, Weiming Zhang, Xinlu Gui, and Bin Yang.
\newblock A novel reversible data hiding scheme based on two-dimensional
  difference-histogram modification.
\newblock {\em IEEE Transactions on Information Forensics and Security},
  8(7):1091--1100, 2013.

\bibitem{puech2008reversible}
William Puech, Marc Chaumont, and Olivier Strauss.
\newblock A reversible data hiding method for encrypted images.
\newblock In {\em Security, Forensics, Steganography, and Watermarking of
  Multimedia Contents X}, volume 6819, page 68191E. International Society for
  Optics and Photonics, 2008.

\bibitem{Zhou2016Secure}
Jiantao Zhou, Weiwei Sun, Li~Dong, Xianming Liu, Oscar~C. Au, and Yuanyang
  Tang.
\newblock Secure reversible image data hiding over encrypted domain via key
  modulation.
\newblock {\em IEEE Transactions on Circuits and Systems for Video Technology},
  26(3):441--452, 2016.

\bibitem{Qian2014Reversible}
Zhenxing Qian, Xinpeng Zhang, and Shuozhong Wang.
\newblock Reversible data hiding in encrypted jpeg bitstream.
\newblock {\em IEEE Transactions on Multimedia}, 16(5):1486--1491, 2014.

\bibitem{Ma2013firstRRBE}
Kede Ma, Weiming Zhang, Xianfeng Zhao, Nenghai Yu, and Fenghua Li.
\newblock Reversible data hiding in encrypted images by reserving room before
  encryption.
\newblock {\em IEEE Transactions on Information Forensics and Security},
  8(3):553--562, 2013.

\bibitem{Xin2017Separable}
Xin Liao, Kaide Li, and Jiaojiao Yin.
\newblock Separable data hiding in encrypted image based on compressive sensing
  and discrete fourier transform.
\newblock {\em Multimedia Tools and Applications}, 76(20):20739--20753, 2017.

\bibitem{Zhang2012Separable}
Xinpeng Zhang.
\newblock Separable reversible data hiding in encrypted image.
\newblock {\em IEEE Transactions on Information Forensics and Security},
  7(2):826--832, 2012.

\bibitem{ShuangPBTL}
Shuang Yi and Yicong Zhou.
\newblock Separable and reversible data hiding in encrypted images using
  parametric binary tree labeling.
\newblock {\em IEEE Transactions on Multimedia}, 21(1):51--64, 2019.

\bibitem{1MSB}
Pauline Puteaux and William Puech.
\newblock An efficient msb prediction-based method for high-capacity reversible
  data hiding in encrypted images.
\newblock {\em IEEE Transactions on Information Forensics and Security},
  13(7):1670--1681, 2018.

\bibitem{puyang2018reversible}
Yi~Puyang, Zhaoxia Yin, and Zhenxing Qian.
\newblock Reversible data hiding in encrypted images with two-msb prediction.
\newblock In {\em 2018 IEEE International Workshop on Information Forensics and
  Security (WIFS)}, pages 1--7. IEEE, 2018.

\bibitem{ccchang}
Kaimeng Chen and ChinChen Chang.
\newblock High-capacity reversible data hiding in encrypted images based on
  extended run-length coding and block-based msb plane rearrangement.
\newblock {\em Journal of Visual Communication and Image Representation},
  58(2019):334--344, 2019.

\bibitem{schaefer2003ucid}
Gerald Schaefer and Michal Stich.
\newblock Ucid: An uncompressed color image database.
\newblock In {\em Storage and Retrieval Methods and Applications for Multimedia
  2004}, volume 5307, pages 472--481. International Society for Optics and
  Photonics, 2003.

\bibitem{boss2011}
Patrick Bas, Tom{\'a}{\v{s}} Filler, and Tom{\'a}{\v{s}} Pevn{\`y}.
\newblock Break our steganographic system: the ins and outs of organizing boss.
\newblock In {\em International workshop on information hiding}, pages 59--70.
  Springer, 2011.

\bibitem{2017BOWS-2}
P.~Bas and T.~Furon.
\newblock Image database of bows-2.
\newblock {\em http://bows2.ec-lille.fr/}, 2017.

\end{thebibliography}

\vspace{-6ex}
\begin{IEEEbiography}[{\includegraphics[width=1in,height=1.25in,clip,keepaspectratio]{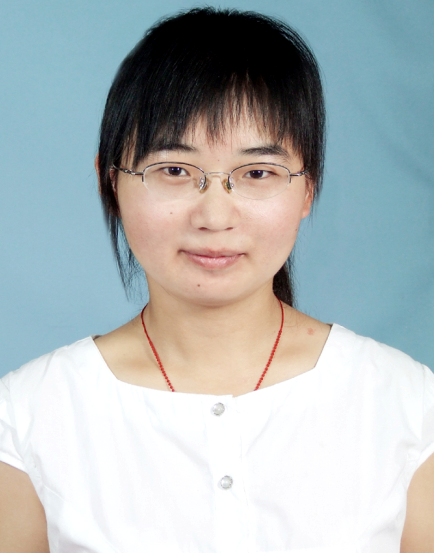}}]{Youqing Wu}
received her B.E. and M.E. in computer science and technology from Anhui University in 2006 and 2009 respectively. She is currently working in School of Computer Science and Technology at Hefei Normal University. Her research interests include Information Hiding and Multimedia Security.
\end{IEEEbiography}

\vspace{-6ex}
\begin{IEEEbiography}[{\includegraphics[width=1in,height=1.25in,clip,keepaspectratio]{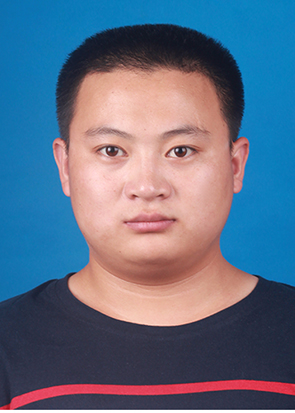}}]{Youzhi Xiang}
received his bachelor degree in computer science and technology in 2017 and now is a master student in the School of Computer Science and Technology, Anhui University. His current research interests include reversible data hiding in encrypted images.
\end{IEEEbiography}

\vspace{-6ex}
\begin{IEEEbiography}[{\includegraphics[width=1in,height=1.25in,clip,keepaspectratio]{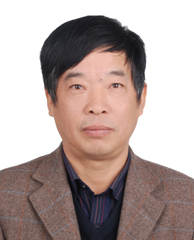}}]{Yutang Guo}
 received his B.Eng., M.Eng. \& Ph.D. degrees from Anhui university in 1987, 1990 and 2009 respectively. He is currently a full professor and a master’s tutor in School of Computer Science and Technology at Hefei Normal University. His research interests include image processing and pattern recognition.
\end{IEEEbiography}

\vspace{-6ex}
\begin{IEEEbiography}[{\includegraphics[width=1in,height=1.25in,clip,keepaspectratio]{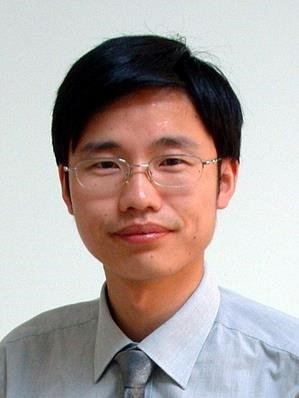}}]{Jin Tang}
 received the B.Eng. degree in automation in 1999, and the Ph.D. degree in computer science in 2007 from Anhui University, Hefei, China. Since 2009, he has been a professor at the School of Computer Science and Technology at the Anhui University. His research interests include image processing, pattern recognition, machine learning and computer vision.
\end{IEEEbiography}

\vspace{-6ex}
\begin{IEEEbiography}[{\includegraphics[width=1in,height=1.25in,clip,keepaspectratio]{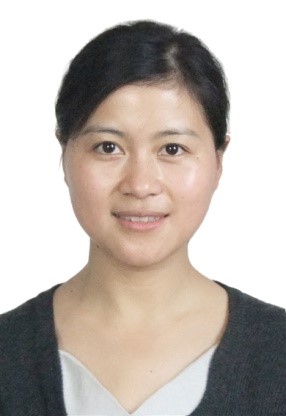}}]{Zhaoxia Yin}
received her B.Sc., M.E. \& Ph.D. from Anhui University in 2005, 2010 and 2014 respectively. She is an IEEE/ACM/CCF member, CSIG senior member and the Associate Chair of the academic committee of CCF YOCSEF Hefei 2016--2017. She is currently working as an Associate Professor and a Ph.D advisor in School of Computer Science and Technology at Anhui University. She is also the Principal Investigator of two NSFC Projects. Her primary research focus including Information Hiding, Multimedia Security and she has published many SCI/EI indexed papers in journals, edited books and refereed conferences.
\end{IEEEbiography}

\end{document}